\documentclass[12pt,a4paper]{article}

\usepackage{epsfig}
\usepackage{amsmath,amsfonts,amssymb}
\usepackage{cite}
\usepackage{colortbl}
\usepackage{pifont}

\providecommand{\openone}{\leavevmode\hbox{\small1\kern-3.8pt\normalsize1}}

\newcommand{\gm}{\gamma^\mu}
\newcommand{\Wm}{W_{\mu}}
\newcommand{\Zm}{Z_{\mu}}
\newcommand{\ts}{$(T)$}
\newcommand{\bs}{$(B)$}
\newcommand{\xt}{$(X\,T)$}
\newcommand{\tb}{$(T\,B)$}
\newcommand{\by}{$(B\,Y)$}
\newcommand{\xtb}{$(X\,T\,B)$}
\newcommand{\tby}{$(T\,B\,Y)$}

\newcommand{\slx}{s_L}
\newcommand{\slu}{s_L^u}
\newcommand{\sld}{s_L^d}
\newcommand{\clx}{c_L}
\newcommand{\clu}{c_L^u}
\newcommand{\cld}{c_L^d}
\newcommand{\srx}{s_R}
\newcommand{\sru}{s_R^u}
\newcommand{\srd}{s_R^d}
\newcommand{\crx}{c_R}
\newcommand{\cru}{c_R^u}
\newcommand{\crd}{c_R^d}

\newcommand{\sqt}{\sqrt{2}}

\newcommand{\RE}{\text{Re}}

\begin{document}

\begin{center}
\begin{Large}
{\bf A handbook of vector-like quarks: \\[1mm]
mixing and single production}
\end{Large}

\vspace{0.5cm}
J.~A.~Aguilar--Saavedra$^{a,b,c}$, R.~Benbrik$^{c,d}$, S.~Heinemeyer$^c$, M.~P\'erez-Victoria$^a$\\[1mm]
\begin{small}
{\it $^a$ Departamento de F\'{\i}sica Te\'orica y del Cosmos, 
Universidad de Granada, Granada, Spain} \\
{\it $^b$Departamento de Fisica, Universidade de Coimbra,  Coimbra, Portugal} \\
{\it $^c$ Instituto de F\'{\i}sica de Cantabria (CSIC-UC), Santander, Spain} \\
{\it $^d$ Faculte Polydisciplinaire de Safi, Sidi Bouzid B.P 4162, 46000 Safi, Morocco}
\end{small}
\end{center}

\begin{abstract}
We obtain constraints on the mixing of vector-like quarks coupling predominantly to the third generation. We consider all (seven) relevant types of vector-like quarks, individually. The constraints are derived from oblique corrections and $Z \to b \bar b$ measurements at LEP and SLC. We investigate the implications of these constraints on LHC phenomenology, concerning the decays of the heavy quarks and their single production. We also explore indirect effects of heavy quark mixing in top and bottom couplings. A remarkable effect is the possibility of explaining the anomalous forward-backward asymmetry in $Z \to b \bar b$ at LEP with a hypercharge $-5/6$ doublet. We also study the impact of the new quarks on single Higgs production at the LHC and Higgs decay.
\end{abstract}
\section{Introduction}

Vector-like quarks are hypothetical spin 1/2 particles that transform as triplets under the colour gauge group and whose left- and right-handed components have the same colour and electroweak quantum numbers. These new particles are receiving a lot of attention for several reasons. To start with, they are the simplest example of coloured fermions still allowed by experimental data. Indeed, extra quarks with chiral couplings, such as fourth generation quarks, are now excluded~\cite{Djouadi:2012ae} by the recent measurements of Higgs-mediated cross sections~\cite{ATLAS:2013sla,CMS:yva}, when combined with direct searches at the Large Hadron Collider (LHC)~\cite{ATLAS:2012qe,Chatrchyan:2012vu}.\footnote{A fourth generation is independently excluded by electroweak precision tests \cite{Eberhardt:2012gv}.} Vector-like quarks, on the other hand, do not receive their masses from Yukawa couplings to a Higgs doublet, and are consistent with existing Higgs data. Secondly, they can mix with the Standard Model (SM) quarks and thereby modify their couplings to the $Z$, $W$ and Higgs boson. Indeed, the addition of vector-like quarks to the SM is the simplest way of breaking the Glashow-Iliopoulos-Maiani~\cite{Glashow:1970gm} mechanism, giving rise for example to tree-level flavour-changing neutral currents~\cite{delAguila:1982fs,Branco:1986my} and potentially striking new effects in low energy physics, none of which have been observed, however.
In this respect, new vector-like quarks also introduce new sources of CP violation~\cite{Nir:1990yq,Branco:1992wr,delAguila:1997vn,Barenboim:1997qx}, as it typically occurs in most SM extensions. In the third place, they can be analysed in a model-independent approach in terms of just a few free parameters. Finally, vector-like quarks at the TeV scale are strongly motivated by at least two theoretical ideas, which are often put together: they are required if the Higgs is a pseudo-Goldstone boson to induce electroweak breaking and explain the observed lightness of the Higgs~\cite{Perelstein:2003wd,Contino:2006qr,Matsedonskyi:2012ym}, and they emerge as fermion resonances in the partial-compositeness theory of flavour~\cite{Kaplan:1991dc,Contino:2006nn}.  Due to the large Yukawa coupling of the top quark, both mechanisms give rise to a sizable mixing of the new quarks with the third family of SM quarks, hence the name ``top partners'', often used in this context. Of course, vector-like quarks do arise in all explicit models that implement these ideas, such as little Higgs and composite Higgs models, or their holographic versions. They also appear in any model with quarks propagating in the bulk of extra dimensions and in grand unified and string theories based on the group $E_6$~\cite{Hewett:1988xc}, although in this case they are not guaranteed to be near the TeV scale.

Extra heavy quarks can be pair produced at hadron colliders through their gauge couplings to gluons, with a strength given by the strong coupling constant, $g_s$. Unlike fermions in other colour representations, they subsequently decay into SM particles, namely ordinary quarks plus a Higgs or a gauge boson, with branching ratios that are mostly determined by their gauge quantum numbers. These decays occur through the mixing of the new quarks with the SM ones. The very same mixing gives rise to two other important effects: it produces a modification of the couplings of the SM quarks (more precisely, of the lighter eigenstates), and it opens up the possibility of single production of the new quarks (the heavier eigenstates), which becomes the dominant production mechanism for high enough masses.

In this paper, we analyse the observable consequences of the mixing between SM and extra vector-like quarks, with emphasis on the connection between them. We find limits on the mixings from electroweak precision data at the $Z$ pole, and use them to extract the allowed values for the mass splittings of the new quarks, their decay branching ratios, the rates of single production and the deviations in the couplings of the known quarks. We follow a model-independent approach by studying all the gauge-covariant fermion multiplets that can mix with the SM quarks via renormalisable couplings, with the implicit assumption that the scalar sector comprises only $\text{SU}(2)_L$ doublets, as is the case of the SM. The possible multiplets have been classified in~\cite{delAguila:2000aa}. It turns out that there are only seven possibilities, so a comprehensive study is a relatively simple task. This systematic approach has already been applied to pair production of vector-like quarks in~\cite{AguilarSaavedra:2009es}. (See also~\cite{AguilarSaavedra:2005pv,AguilarSaavedra:2006gv,AguilarSaavedra:2006gw,Azatov:2012rj,Harigaya:2012ir}.)

In order to simplify the analysis, we assume that the vector-like quarks only couple to the quarks of the third family (for a more detailed discussion see~\cite{Aguilar-Saavedra:2013wba}). As we have mentioned above, this pattern of mixing is well motivated by the large Yukawa coupling of the top quark, which suggests a close connection of the top quark (and the left-handed component of the bottom quark) with any new physics related to electroweak symmetry breaking or to the fermion mass hierarchy. From an experimental point of view, predominant mixing with the third generation is favoured as well. In the down sector, it helps in avoiding flavour problems, thanks to the hierarchical structure of the CKM matrix, although exclusive mixing with the $d$ or $s$ quarks is also allowed. In the up sector the mixing with the $u$ or $c$ quarks is subdominant if only one vector-like multiplet is included, but it leads to interesting flavour-changing neutral top interactions~\cite{delAguila:1998tp}.\footnote{We also note that a cancellation of the effects of different extra quarks might allow a significant mixing with the lighter generations~\cite{Atre:2008iu}. The phenomenological consequences for LHC searches have been explored in~\cite{Atre:2011ae}.}

Electroweak precision observables are sensitive to the sum of the different possible new physics contributions. For this reason, to extract precise limits on new particles it is always necessary to make some assumption that restricts new effects on electroweak precision data. We make an unbiased choice based on minimality: we consider extensions of the SM with all possible vector-like quark multiplets, but only one at a time, and further assume that no additional new physics modifies the electroweak observables. The bounds on the mixings we derive in this fashion are conservative for most new physics scenarios. Indeed, due to the large range of electroweak observables and the combination of tree-level and loop contributions, additional multiplets or other new physics effects typically increase the $\chi^2$ of the fits \cite{delAguila:2011yd}. However, one should always keep in mind the existence of models that are constructed to enforce cancellations that relax the bounds, possibly with the aid of symmetries. Well motivated examples are given by realistic composite Higgs models, based on~\cite{Agashe:2004rs}, which contain several vector-like multiplets and incorporate a custodial protection at the tree level of the T parameter~\cite{Peskin:1990zt} and the ratio $R_b$ of the partial width for $Z \to b \bar b$ over the total hadronic $Z$ width. Such models~\cite{Agashe:2003zs,Agashe:2006at} are not covered by the analysis in this paper. They require dedicated examinations of their electroweak constraints (see e.g.~\cite{Anastasiou:2009rv}) and their LHC implications (see~\cite{Contino:2008hi,Mrazek:2009yu,DeSimone:2012fs}).

We anticipate that, for most multiplets, the electroweak constraints that we derive with our assumptions are quite tight, and preclude a clear observation of deviations in the couplings of the top quark at the LHC and the International Linear Collider (ILC). On the other hand, we find that a quark doublet with hypercharge $-5/6$ has relatively weak bounds and can actually improve significantly the electroweak fit by reconciling, with just one free parameter, the predictions for $R_b$ and the forward-backward (FB) asymmetry in $e^+ e^- \to Z \to b \bar b$, $A_\mathrm{FB}^b$, with their observed values at the Large Electron Positron collider (LEP) and the Stanford Linear Collider (SLC)~\cite{ALEPH:2005ab}. This remarkably simple explanation of the long-standing $A_\mathrm{FB}^b$ anomaly was originally proposed in \cite{Choudhury:2001hs}.\footnote{In that paper, a vector-like singlet was also added to keep $R_b$ close to its SM value (see~\cite{Kumar:2010vx} for a dedicated analysis of the LHC phenomenology of this model). Here we explore the simplest possibility of fitting the electroweak data with only one multiplet.} It predicts a large single production rate of an extra quark of electric charge $-4/3$, with a visible signal at the LHC.

The paper is organised as follows. In Section~\ref{sec:mix} we introduce the seven vector-like quark multiplets and describe their mixing with the third family of SM quarks. In Section~\ref{sec:lim} we obtain limits on these mixings and the masses of the new quarks from the relevant electroweak precision data, namely oblique parameters and $Z\to b\bar{b}$ observables. Section~\ref{sec:higgs} contains a brief discussion on the possible effects in Higgs production and decay. In Section~\ref{sec:dec} we study the allowed splittings of masses and the allowed branching ratios for the decay of all the different extra quarks. In Section~\ref{sec:xsec} we analyse the allowed single production cross sections at the LHC. In Section~\ref{sec:top} we discuss the allowed deviations of top couplings and the expectations for measurements at the LHC and ILC. Section~\ref{sec:fit} is devoted to the vector-like-quark explanation of the $A_\mathrm{FB}^b$ LEP anomaly and its observable consequences at the LHC. We conclude in Section~\ref{sec:sum}. Finally, Appendix~\ref{sec:a} collects the analytical expressions for all the couplings of light and heavy quarks to the gauge bosons and the Higgs, and Appendix~\ref{sec:b} the partial widths for the different heavy quark decay modes.


\section{Mixing with vector-like quarks}
\label{sec:mix}

If the scalar sector only includes $\text{SU}(2)_L$ doublets, as is the case of the SM, new vector-like quarks coupling to the SM ones with renormalisable couplings can only appear in seven gauge-covariant multiplets with definite $\text{SU}(3)_C \times \text{SU}(2)_L \times \text{U}(1)_Y$ quantum numbers~\cite{delAguila:2000aa}:
\begin{align}
& T_{L,R}^0 \,, \quad B_{L,R}^0 && \text{(singlets)} \,, \notag \\
& (X\,T^0)_{L,R} \,, \quad (T^0\,B^0)_{L,R} \,, \quad (B^0\,Y)_{L,R} && \text{(doublets)} \,, \notag \\
& (X\,T^0\,B^0)_{L,R} \,, \quad (T^0\,B^0\,Y)_{L,R}  && \text{(triplets)} \,.
\end{align}
We use in this section a zero superscript on weak eigenstates to distinguish them from mass eigenstates; this superscript will be omitted when it is clear from the context. The new fields $T^0$, $B^0$ have electric charges $2/3$ and $-1/3$, respectively.
Note that some of the multiplets include quarks $X$ of electric charge $5/3$, and $Y$ with charge $-4/3$, in which case the weak and mass eigenstates coincide as long as only one such multiplet is present. We will actually restrict ourselves to extensions of the SM with only one extra multiplet, as explained in the introduction.

When new fields $T_{L,R}^0$ of charge $2/3$ and non-standard isospin assignments are added to the SM, the resulting physical up-type quark mass eigenstates $u,c,t,T$ may in general contain non-zero $T_{L,R}^0$ components, leading for example to a deviation in their couplings to the $Z$ boson. Constraints on these deviations for the up and charm quarks result from atomic parity violation experiments and the measurement of $R_c$ at LEP~\cite{Beringer:1900zz}, and are far stronger than for the top quark~\cite{AguilarSaavedra:2002kr}.\footnote{This statement applies not only to $T$ singlets~\cite{AguilarSaavedra:2002kr} but to all multiplets with $T$ quarks, since in all cases the deviations in the $Z$ couplings are given by the square of a mixing angle.}
 So, it is very reasonable to assume that the only the top quark has sizeable $T_{L,R}^0$ components (or, in other words, only the top quark ``mixes'' with $T$). In this case, the relation between charge $2/3$ weak and mass eigenstates can be parameterised by two $2 \times 2$ unitary matrices $U_{L,R}^u$,
\begin{equation}
\left(\! \begin{array}{c} t_{L,R} \\ T_{L,R} \end{array} \!\right) =
U_{L,R}^u \left(\! \begin{array}{c} t^0_{L,R} \\ T^0_{L,R} \end{array} \!\right)
= \left(\! \begin{array}{cc} \cos \theta_{L,R}^u & -\sin \theta_{L,R}^u e^{i \phi_u} \\ \sin \theta_{L,R}^u e^{-i \phi_u} & \cos \theta_{L,R}^u \end{array}
\!\right)
\left(\! \begin{array}{c} t^0_{L,R} \\ T^0_{L,R} \end{array} \!\right) \,.
\label{ec:mixu}
\end{equation}
In the down sector, the addition of new fields $B_{L,R}^0$ of charge $-1/3$ results in four mass eigenstates $d,s,b,B$. In contrast with the up sector, the measurement of $R_b$ at LEP sets constraints on the $b$ mixing with the new fields that are stronger than for mixing with the lighter quarks $d,s$~\cite{AguilarSaavedra:2002kr}. (An exception to this statement is discussed in Section~\ref{sec:fit}.) However, one still expects dominant mixing with the $b$ quark given the usual Yukawa coupling hierarchy in the mass matrices. This is the case, for instance, in models with fermion partial compositeness. We will then assume dominant $b-B$ mixing, parameterised by two $2 \times 2$ unitary matrices $U_{L,R}^d$,
\begin{equation}
\left(\! \begin{array}{c} b_{L,R} \\ B_{L,R} \end{array} \!\right)
= U_{L,R}^d \left(\! \begin{array}{c} b^0_{L,R} \\ B^0_{L,R} \end{array} \!\right)
= \left(\! \begin{array}{cc} \cos \theta_{L,R}^d & -\sin \theta_{L,R}^d e^{i \phi_d} \\ \sin \theta_{L,R}^d e^{-i \phi_d} & \cos \theta_{L,R}^d \end{array}
\!\right)
\left(\! \begin{array}{c} b^0_{L,R} \\ B^0_{L,R} \end{array} \!\right) \,.
\label{ec:mixd}
\end{equation}
The Lagrangian for the third generation and heavy quarks in the mass eigenstate basis is given in Appendix~\ref{sec:a}. To ease the notation, we have dropped the superscripts $u$ ($d$) of the angles $\theta^u_{L,R}$ ($\theta^d_{L,R}$) in the models where the mixing occurs only in the up (down) sector. Additionally, we use the shorthands $s_{L,R}^{u,d} \equiv \sin \theta_{L,R}^{u,d}$, $c_{L,R}^{u,d} \equiv \cos \theta_{L,R}^{u,d}$, etc.
This Lagrangian contains all the phenomenologically relevant information:
\begin{itemize}
\item[(i)] the modifications of the SM couplings that might show indirect effects of new quarks can be found in the terms that do not contain heavy quark fields;
\item[(ii)] the terms relevant for LHC phenomenology ---heavy quark production and decay--- are those involving a heavy and a light quark;
\item[(iii)] terms with two heavy quarks are relevant for their contribution to oblique corrections.
\end{itemize}
The unitary matrices $U_{L,R}^u$ in Eq.~(\ref{ec:mixu}) and $U_{L,R}^d$ in Eq.~(\ref{ec:mixd}) are determined by the condition that the mass matrices in the mass eigenstate basis are diagonal. In the weak eigenstate basis, the third generation and heavy quark mass terms are
\begin{eqnarray}
\mathcal{L}_\text{mass} & = & - \left(\! \begin{array}{cc} \bar t_L^0 & \bar T_L^0 \end{array} \!\right)
\left(\! \begin{array}{cc} y_{33}^u \frac{v}{\sqrt 2} & y_{34}^u \frac{v}{\sqrt 2} \\ y_{43}^u \frac{v}{\sqrt 2} & M^0 \end{array} \!\right)
\left(\! \begin{array}{c} t^0_R \\ T^0_R \end{array}
\!\right) \notag \\
& & - \left(\! \begin{array}{cc} \bar b_L^0 & \bar B_L^0 \end{array} \!\right)
\left(\! \begin{array}{cc} y_{33}^d \frac{v}{\sqrt 2} & y_{34}^d \frac{v}{\sqrt 2} \\ y_{43}^d \frac{v}{\sqrt 2} & M^0 \end{array} \!\right)
\left(\! \begin{array}{c} b^0_R \\ B^0_R \end{array}
\!\right) +\text{H.c.} \,,
\label{ec:Lmass}
\end{eqnarray}
with $y_{ij}^q$, $q=u,d$, Yukawa couplings, $v=246$ GeV the Higgs vacuum expectation value (VEV) and $M^0$ a bare mass
term.\footnote{As pointed out in the introduction, this bare mass term is not related to the Higgs mechanism. It is gauge-invariant and can appear as a bare mass term in the Lagrangian, or it can be generated by a Yukawa coupling to a scalar singlet that acquires a VEV $v' \gg v$.} Then, the mixing matrices are determined by
\begin{equation}
U_L^q \, \mathcal{M}^q \, (U_R^q)^\dagger = \mathcal{M}^q_\text{diag} \,, 
\label{ec:diag}
\end{equation}
with $\mathcal{M}^q$ the two mass matrices in Eq.~(\ref{ec:Lmass}) and $\mathcal{M}^q_\text{diag}$ the diagonalised ones. These general equations are simplified in some particular cases. In the multiplets where either $T$ or $B$ quarks are absent, the corresponding $2 \times 2$ mass matrix reduces to the SM quark mass term. Notice also that, in multiplets with both $T$ and $B$ quarks, the bare mass term is the same for the up and down sectors. For singlets and triplets one has $y_{43}^q = 0$, whereas for doublets $y_{34}^q=0$. Moreover, for the \xtb\ triplet $y_{34}^d = \sqrt 2 y_{34}^u$, and for the \tby\ triplet, $y_{34}^u = \sqrt 2 y_{34}^d$.\footnote{We write the triplets in the spherical basis. The $\sqrt 2$ factors stem from the relation between the cartesian and spherical coordinates of an irreducible tensor operator of rank 1 (vector).}

The mixing angles in the left- and right-handed sectors are not independent parameters. From the mass matrix bi-unitary diagonalisation in Eq.~(\ref{ec:diag}) one finds (see also~\cite{Fajfer:2013wca})
\begin{eqnarray}
\tan 2 \theta_L^q & = & \frac{\sqrt{2} |y_{34}^q| v M^0}{(M^0)^2-|y_{33}^q|^2 v^2/2 - |y_{34}^q|^2 v^2/2} \quad \text{(singlets, triplets)} \,, \notag \\
\tan 2 \theta_R^q & = & \ \frac{\sqrt{2}  |y_{43}^q| v M^0}{(M^0)^2-|y_{33}^q|^2 v^2/2 - |y_{43}^q|^2 v^2/2} \quad \text{(doublets)} \,, 
\label{ec:angle1}
\end{eqnarray}
with the relations (see also~\cite{Dawson:2012di,Fajfer:2013wca,Atre:2011ae})
\begin{eqnarray}
\tan \theta_R^q & = & \frac{m_q}{m_Q} \tan \theta_L^q \quad \text{(singlets, triplets)} \,, \notag \\
\tan \theta_L^q & = & \frac{m_q}{m_Q} \tan \theta_R^q \quad \text{(doublets)} \,, 
\label{ec:rel-angle1}
\end{eqnarray}
with $(q,m_q,m_Q) = (u,m_t,m_T),(d,m_b,m_B)$, so one of the mixing angles is always dominant, especially in the down sector. In addition, for the triplets the relations between the off-diagonal Yukawa couplings lead to relations between the mixing angles in the up and down sectors,
\begin{eqnarray}
\sin 2\theta_L^d & = & \sqrt{2} \, \frac{m_T^2-m_t^2}{m_B^2-m_b^2} \sin 2 \theta_L^u \quad \quad (X\,T\,B) \,, \notag \\
\sin 2\theta_L^d & = & \frac{1}{\sqrt{2}} \frac{m_T^2-m_t^2}{m_B^2-m_b^2} \sin 2 \theta_L^u \quad \quad (T\,B\,Y) \,.
\end{eqnarray}
Therefore, all multiplets involve a single independent mixing angle parameter, except the \tb\ doublet, which has two. The masses of the heavy quarks deviate from $M^0$ due to the non-zero mixing with the SM quarks, and for doublets and triplets the masses of the different components of the multiplet are related, as described in Section~\ref{sec:split}. Altogether, these relations show that all multiplets except the \tb\ doublet can be parameterised by a mixing angle, a heavy quark mass and a CP-violating phase that enters few couplings and can be ignored for the observables considered in this paper. In the case of the \tb\ doublet there are two independent mixing angles and two CP-violating phases for the up and down sectors.


\section{Limits on mixing}
\label{sec:lim}

The mixing of the top and bottom quark with heavy partners results in new contributions to the oblique parameters $\text{S}$ and $\text{T}$~\cite{Peskin:1990zt}, precisely measured at LEP and SLC.\footnote{Changes in the $\text{U}$ parameter are subleading, as for any new physics at a scale much higher than the mass of the $Z$, while the oblique parameters $\text{Y}$ and $\text{W}$, introduced in~\cite{Barbieri:2004qk} and relevant for LEP~2 observables, are not modified in these extensions of the SM.} The contributions to $\text{S},\text{T}$ in models with arbitrary numbers of $T,B$ singlets and \tb\ doublets were given in~\cite{Lavoura:1992np}, and generalised for arbitrary vector-like quarks in~\cite{Carena:2006bn} for S and \cite{Anastasiou:2009rv} for T. We have computed the contributions of new quarks to $\Delta \text{T} = \text{T}-\text{T}_\text{SM}$ and $\Delta \text{S} = \text{S}-\text{S}_\text{SM}$ using the analytical expressions in these references. For the $T$ singlet and \xt, \tb\ doublets, these calculations have been cross-checked by implementing the models in FeynArts/FormCalc~\cite{Hahn:2000kx,Hahn:1998yk}, which are then used to calculate the gauge boson self-energies. We take the experimental values, for $\Delta\text{U}=0$,
\begin{align}
& \Delta \text{S} = 0.04 \pm 0.07 \,, \notag \\
& \Delta \text{T} = 0.07 \pm 0.08 \,,
\end{align}
with a correlation of 0.88~\cite{Beringer:1900zz}. The largest deviations are found, for all multiplets, in the T parameter. Thus, it is the modification of this quantity which determines the upper limits on mixing angles, as we will see below.

\begin{figure}[p]
\begin{center}
\begin{tabular}{cc}
\includegraphics[height=5.2cm,clip=]{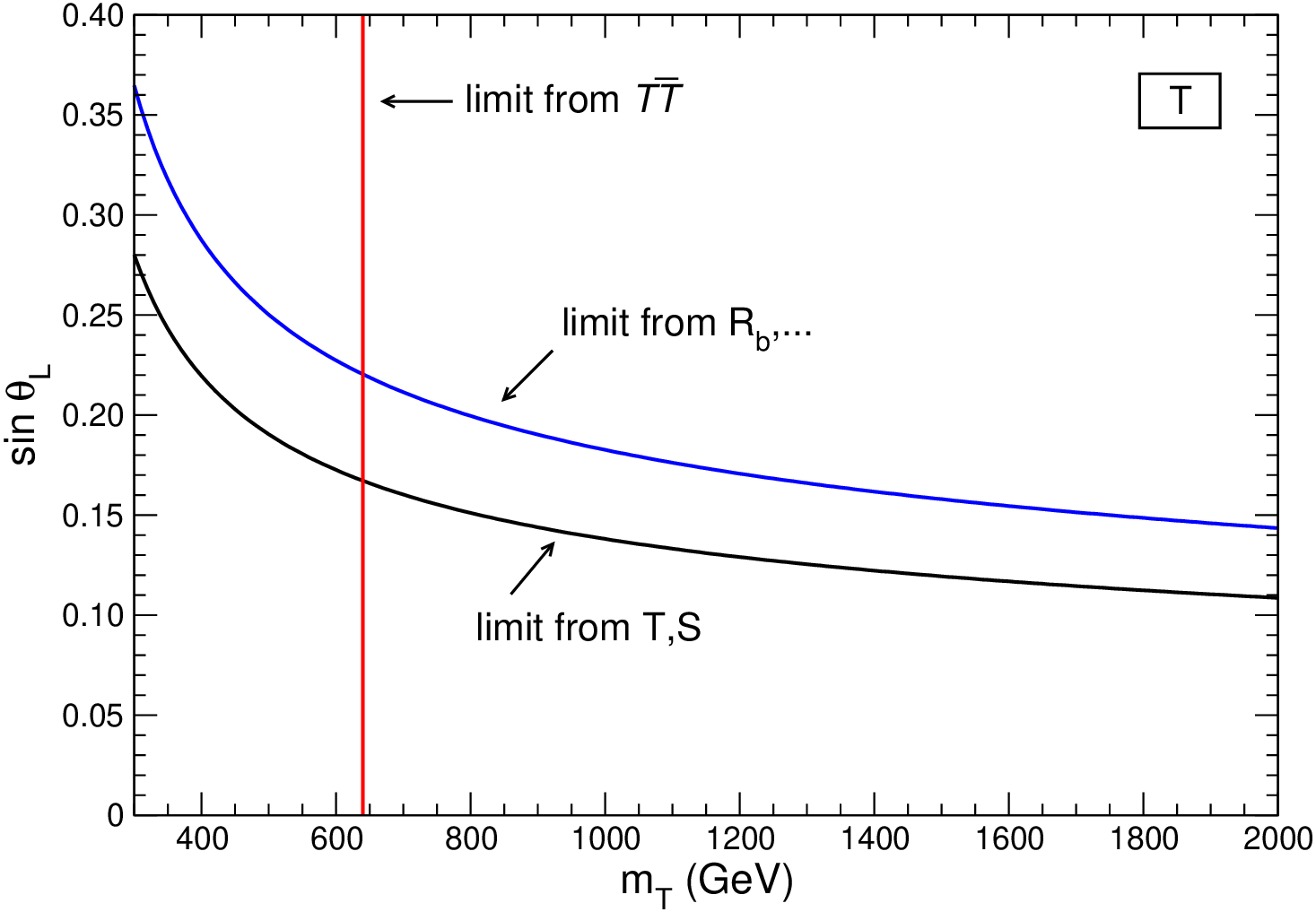} & \includegraphics[height=5.2cm,clip=]{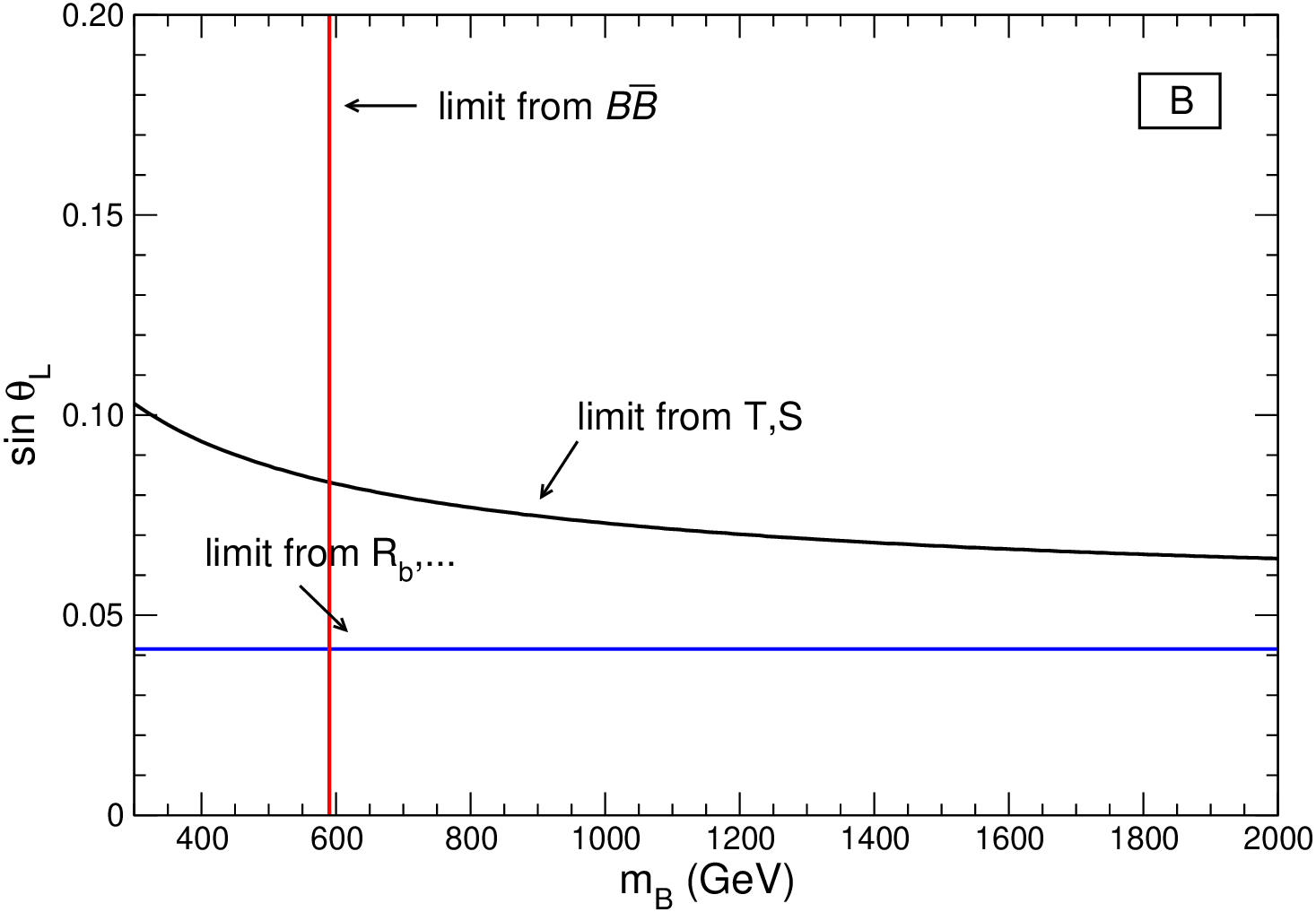} \\
\includegraphics[height=5.2cm,clip=]{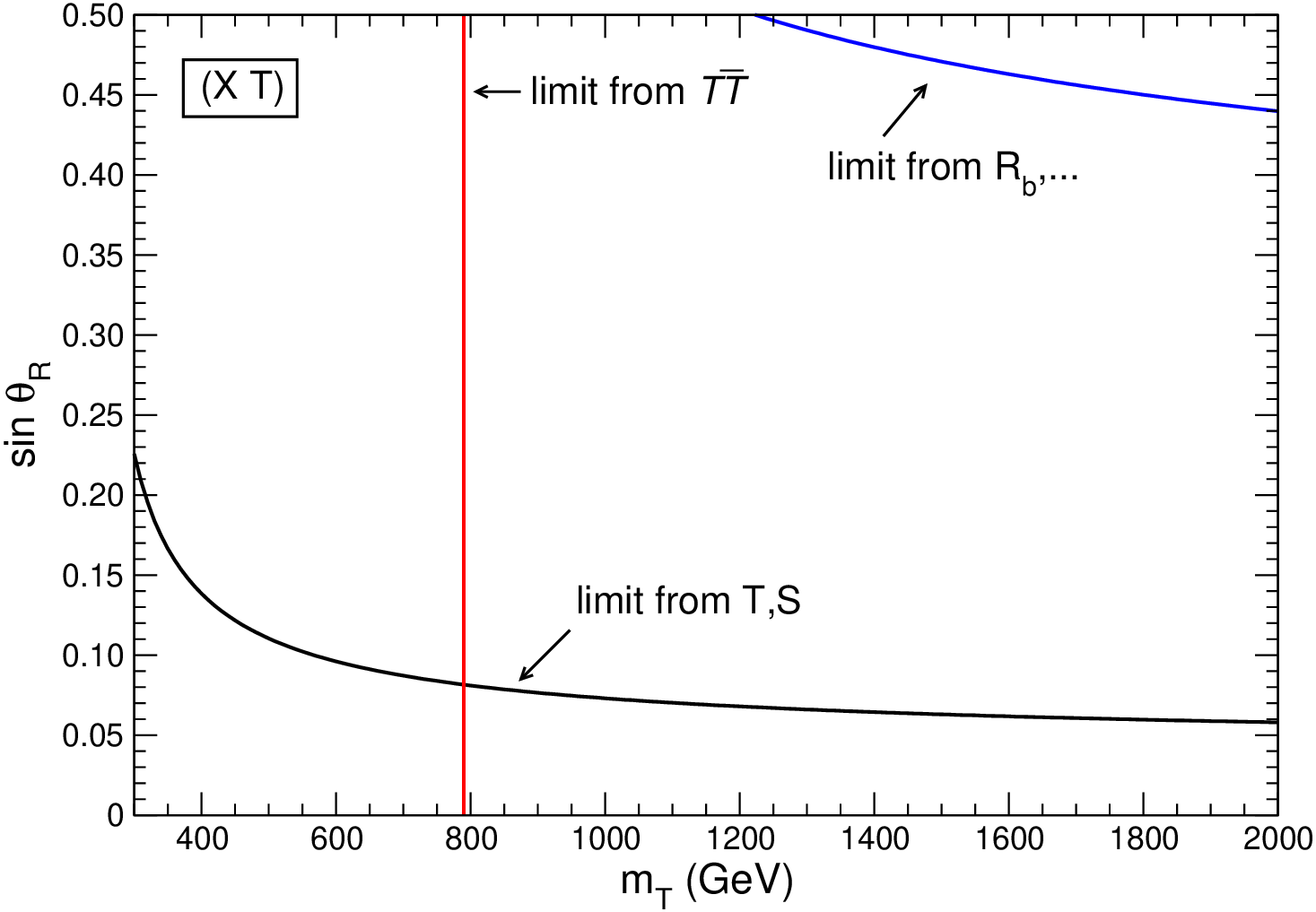} & \includegraphics[height=5.2cm,clip=]{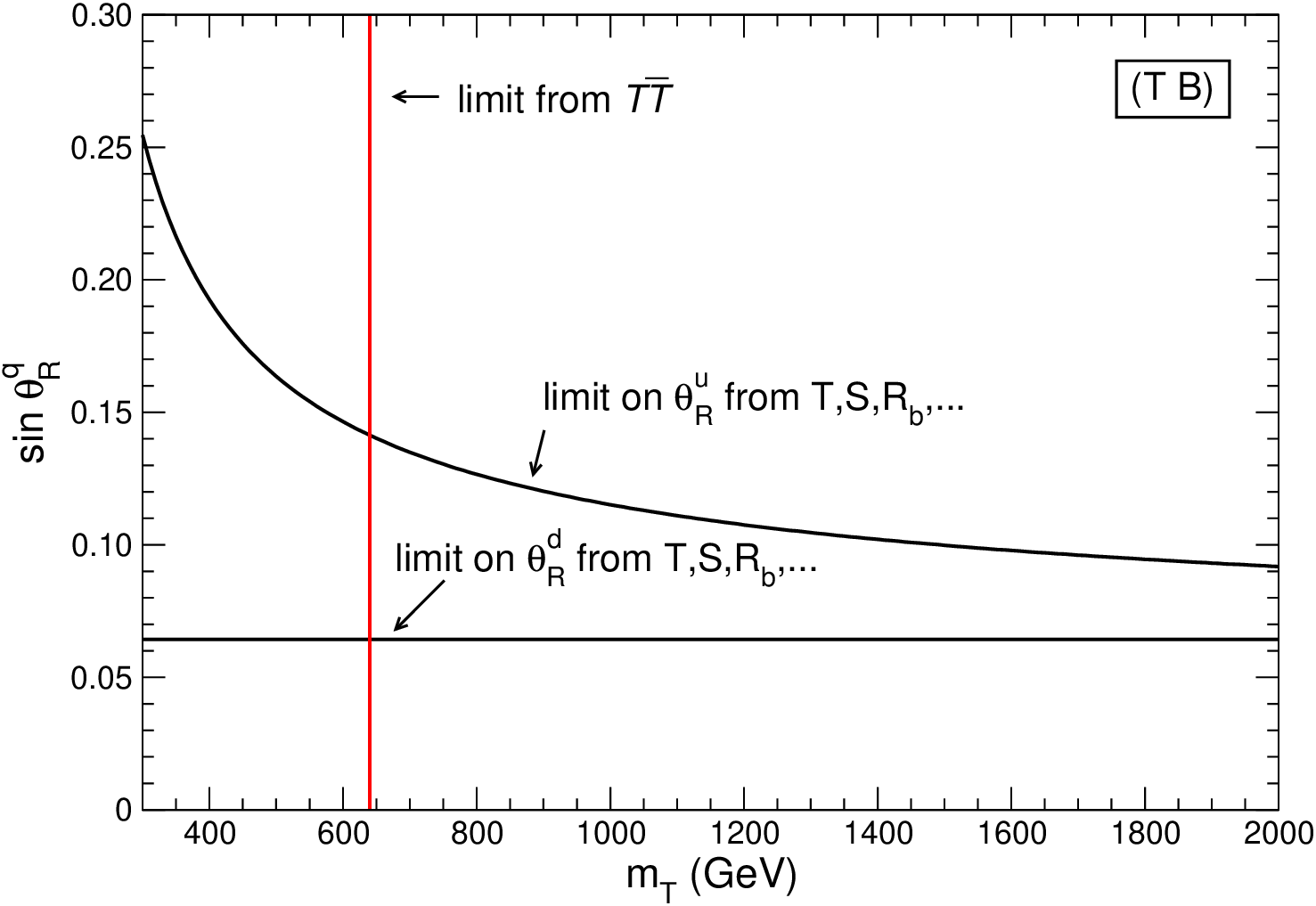} \\
\includegraphics[height=5.2cm,clip=]{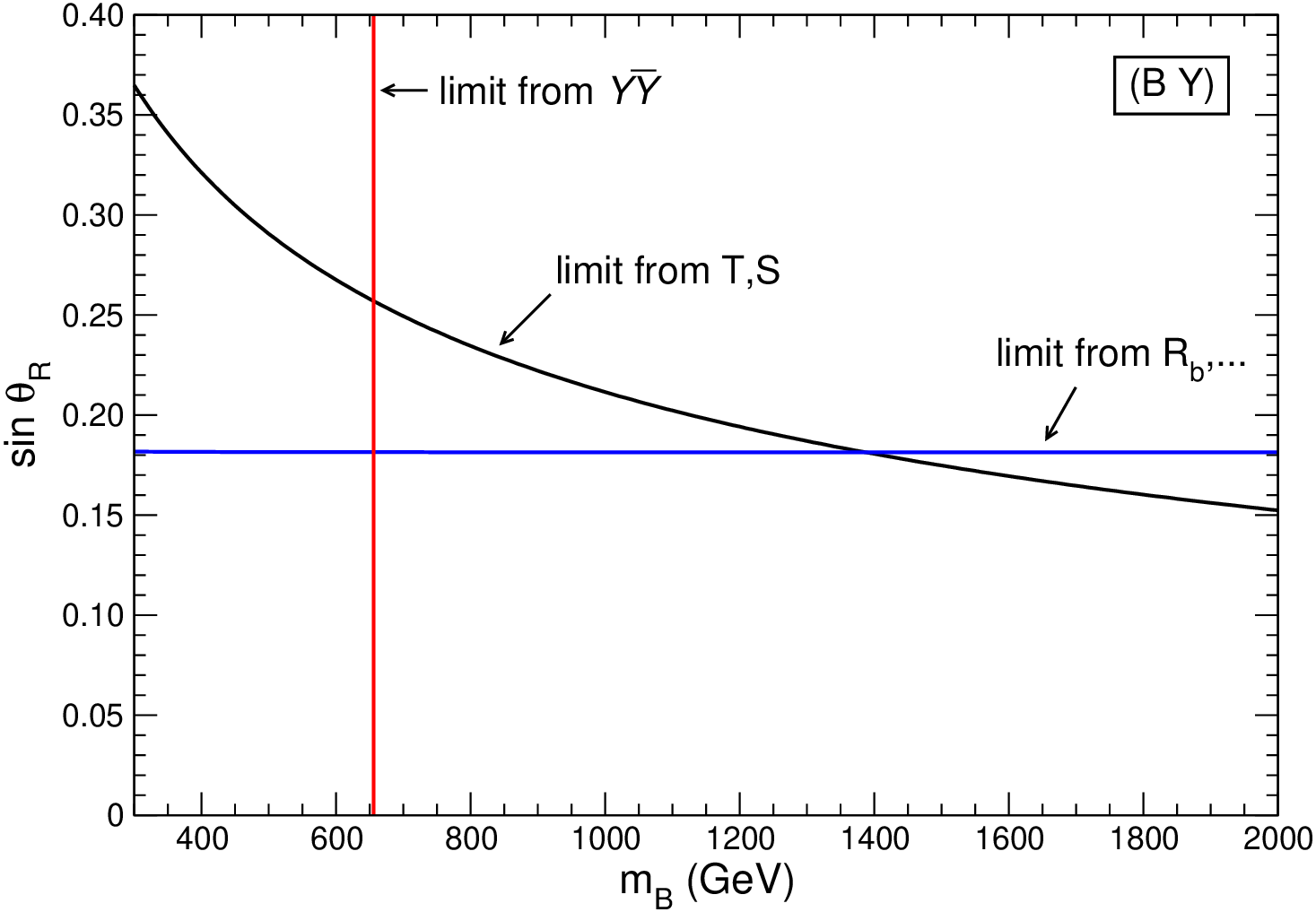} & \includegraphics[height=5.2cm,clip=]{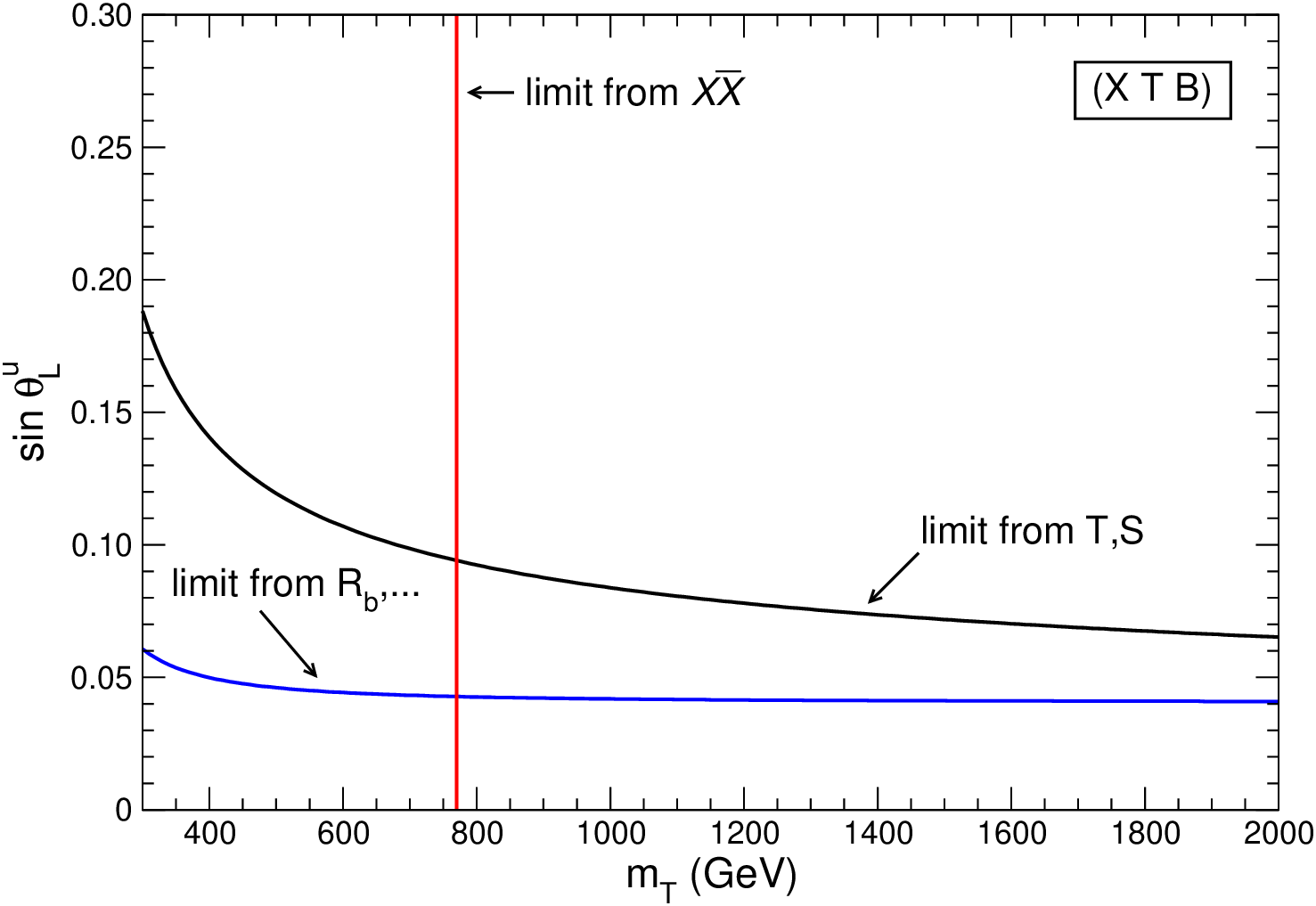} \\
\multicolumn{2}{c}{\includegraphics[height=5.2cm,clip=]{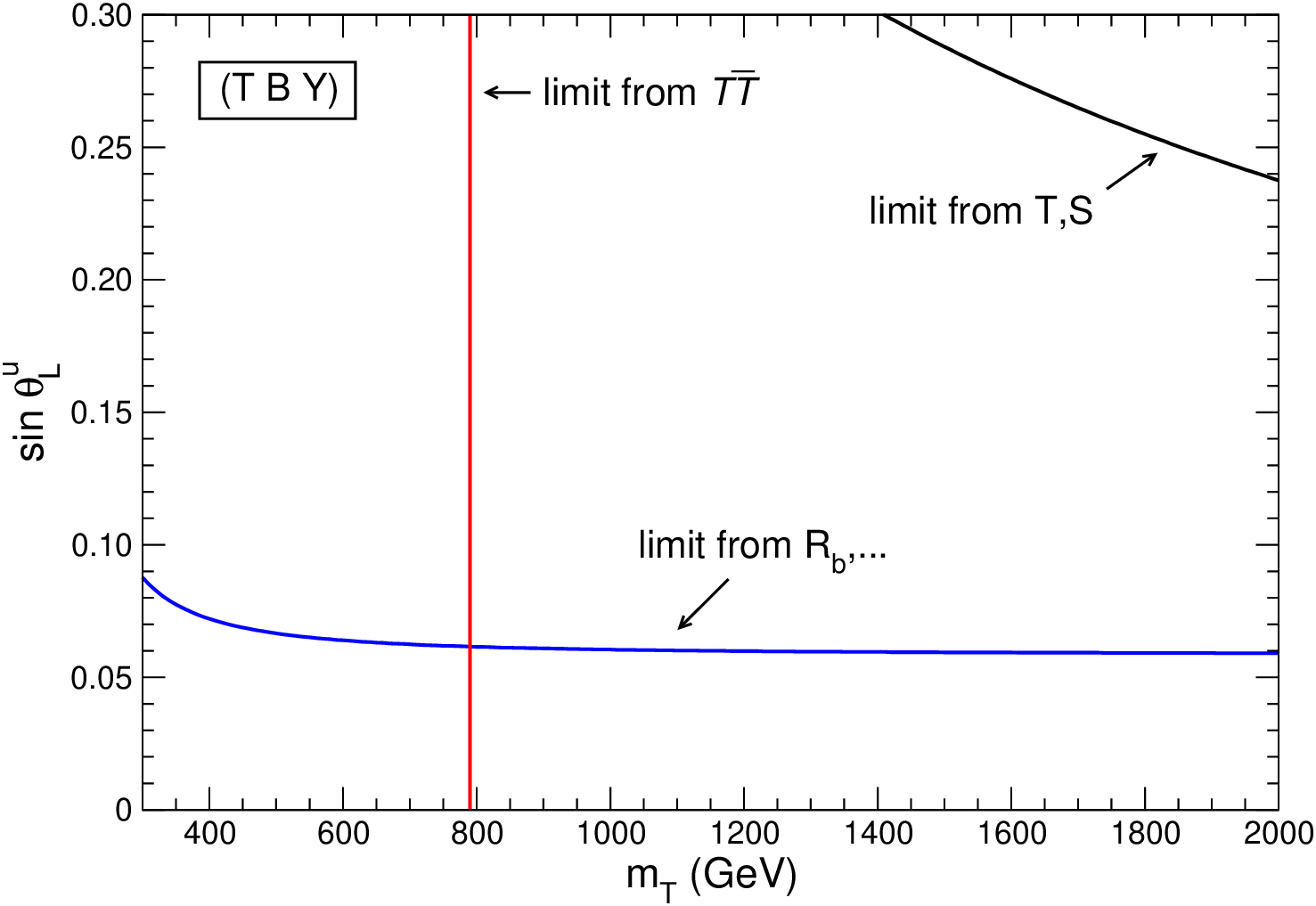}}
\end{tabular}
\caption{Upper limits on the mixing angles for the different multiplets. The current mass limits from direct searches are also indicated by vertical lines.}
\label{fig:lim}
\end{center}
\end{figure}

The mixing of the $b$ quark with a heavy $B$ modifies the $Z b \bar b$ coupling at the tree level, whereas $t-T$ mixing modifies it at the one loop level, via the top correction to the effective $Z b_L b_L$ vertex~\cite{Bernabeu:1987me}. We compute this correction in the presence of vector-like quarks using the analytical expressions in~\cite{Bamert:1996px}. For the SM predictions, we use the values from a fit in~\cite{Beringer:1900zz}:
\begin{eqnarray}
R_b^\text{SM} & = & 0.21576 \,, \notag \\
A_\text{FB}^{b,\text{SM}} & = & 0.1034 \,, \notag \\
A_b^\text{SM} & = & 0.9348 \,, \notag \\
R_c^\text{SM} & = & 0.17227 \,.
\end{eqnarray}
These four observables are affected by modifications of the $Zbb$ vertex, although $R_c$ only indirectly. Writing the effective $Zbb$ vertex as
\begin{equation}
\mathcal{L}_{Zbb} = -\frac{g}{2c_W} \bar b \gm (c_L P_L + c_R P_R) b \Zm \,,
\label{ec:Zbb}
\end{equation}
and assuming small shifts of the effective couplings $\delta c_L$, $\delta c_R$, the deviations in these observables with respect to the SM values can be well approximated by the first-order expressions
\begin{eqnarray}
R_b & = & R_b^\text{SM} \left(1-1.820 \delta c_L + 0.336 \delta c_R \right) \,, \notag \\
A_\text{FB}^b & = & A_\text{FB}^{b,\text{SM}} \left(1- 0.1640 \delta c_L - 0.8877 \delta c_R \right) \,, \notag \\
A_b & = & A_b^\text{SM} \left(1- 0.1640 \delta c_L - 0.8877 \delta c_R \right) \,, \notag \\
R_c & = & R_c^\text{SM} \left(1 + 0.500 \delta c_L - 0.0924 \delta c_R \right) \,.
\end{eqnarray}
As experimental measurements, we take~\cite{ALEPH:2005ab}
\begin{eqnarray}
R_b^\text{exp} & = & 0.21629 \pm 0.00066 \,, \notag \\
A_\text{FB}^{b,\text{exp}} & = & 0.0992 \pm 0.0016 \,, \notag \\
A_b^\text{exp} & = & 0.923 \pm 0.020 \,, \notag \\
R_c^\text{exp} & = & 0.1721 \pm 0.003 \,,
\label{ec:RbX}
\end{eqnarray}
with the correlation matrix
\begin{equation}
\rho = \left( \! \begin{array}{cccc}
1 & -0.10 & -0.08 & -0.18  \\
-0.10 & 1 & 0.06 & 0.04 \\
-0.08 & 0.06 & 1 & 0.04 \\
-0.18 & 0.04 & 0.04 & 1
\end{array}
\! \right) \,,
\end{equation}
keeping the same ordering of the observables as in Eq.~(\ref{ec:RbX}).
For all multiplets, except the \tb\ doublet, the constraints from $\text{T},\text{S}$ and $Z \to b \bar b$ are independent. We thus compute the 95\% confidence level (CL) upper limits on mixing angles from each set of observables.\footnote{We do not combine $\text{T},\text{S}$ and $Z \to b \bar b$ observables in a joint $\chi^2$, but simply require independent agreement with both sets at the 95\% CL. Since in most cases one of the constraints strongly dominates the other, the 95\% CL interpretation is retained.} The results are presented in Fig.~\ref{fig:lim}.%
For multiplets with a $T$ quark, we take as independent parameters $m_T$ and the dominant (unsuppressed) mixing angle in the up sector, see Eqs.~(\ref{ec:rel-angle1}). Otherwise, we take as parameters $m_B$ and the dominant mixing in the down sector. For illustration, we also include vertical lines corresponding to the current lower limits on the heavy quark masses. (These limits depend on the heavy quark decay modes, which are different for the different multiplets, see Section~\ref{sec:dec} for further details and references.)
For multiplets without a $B$ quark, the constraints from $\text{T},\text{S}$ are more restrictive, whereas for multiplets with a $B$ quark ---where tree-level contributions to the $Zbb$ vertex appear--- the converse holds. For the \tb\ doublet, the constraints on $\theta_R^u$ are determined by $\text{T},\text{S}$, but they slightly depend on the value of $\theta_R^d$, which is constrained by $Z \to b \bar b$. We thus impose agreement with the two angles at the 95\% CL, and present the resulting limits on both $\theta_R^u$ and $\theta_R^d$. The mixing for the \by\ doublet is less constrained than what might be expected due to the existing discrepancies between the $Z\to b \bar b$ data and the SM predictions. We examine this case in detail in Section~\ref{sec:fit}.


\section{Contribution to Higgs production and decay}
\label{sec:higgs}

Vector-like quarks enter the loop diagrams in the amplitudes for Higgs production by gluon-gluon fusion and Higgs decay into two photons. However, the minimal extensions considered in this work give small contributions, as it will be explicitly shown below. One reason is that vector-like quarks decouple when their gauge-invariant masses become large, with fixed Yukawa couplings (in which case their mixing with the SM quarks becomes small). Furthermore, in the case of a heavy $T$ mixing with the top quark, the contribution turns out to be much smaller than what decoupling suggests, due to a cancellation between the amplitudes with heavy-quark loops and the effect of modified couplings in the loops with the top. This mechanism has already been shown for a singlet $T$ in~\cite{AguilarSaavedra:2006gw}, and for a $(TB)$ doublet in~\cite{Dawson:2012di}. In the following, we extend it to all the seven multiplets.

Let us consider  the $gg\to H$ (or $H \to gg$) and $H \to \gamma \gamma$ processes in models with one vector-like multiplet. The contribution of all the quarks of the same charge to the two corresponding amplitudes is proportional to
\begin{equation}
F_q=\sum_i Y_{ii} A_{1/2}\left(\frac{M_H^2}{4 m_i^2} \right),
\end{equation}
where the  sum runs over $t,T$ for $q=u$ and over $b,B$ for $q=d$, in the mass-eigenstate basis; the couplings to the Higgs $Y_{ii}$ are defined in Eqs.~(\ref{ec:ll}) and (\ref{ec:HH}) of Appendix~\ref{sec:a}, and the function $A_{1/2}$ is defined, for instance, in~\cite{Djouadi:2005gi}. $A_{1/2}$ approaches the infinite mass value of 4/3 pretty fast for $m_i$ larger than $M_H$, which holds for both $t$ and new heavy states. (The difference between $A_{1/2}(M_H^2/4 m_t^2)$ and the asymptotic value for large quark masses is of only 3\%.) Thus, in the up sector we can approximate
\begin{equation}
F_u \simeq \frac{4}{3} \left( Y_{tt} + Y_{TT} \right) = \frac{4}{3} \,,
\end{equation}
just as in the SM. The reason for this cancellation can be easily identified. Defining the matrix
\begin{equation}
Y^0=\left( \begin{array}{cc} 1&0\\ 0 & 0 \end{array} \right) \,,
\end{equation}
for singlets and triplets the Higgs interactions are given by
\begin{eqnarray}
\mathcal{L}_H & = & - \frac{1}{v} \left(\! \begin{array}{cc} \bar t_L^0 & \bar T_L^0 \end{array} \!\right)
Y^0 \mathcal{M}^u
\left(\! \begin{array}{c} t^0_R \\ T^0_R \end{array}
\!\right) H + \text{H.c.} \notag \\
& = &
- \frac{1}{v} \left(\! \begin{array}{cc} \bar t_L & \bar T_L \end{array} \!\right) U_L^u
Y^0 (U_L^u)^\dagger \mathcal{M}^u_\text{diag}
\left(\! \begin{array}{c} t_R \\ T_R \end{array}
\!\right) H + \text{H.c.} \,,
\end{eqnarray}
and for doublets by
\begin{eqnarray}
\mathcal{L}_H & = & - \frac{1}{v} \left(\! \begin{array}{cc} \bar t_L^0 & \bar T_L^0 \end{array} \!\right)
 \mathcal{M}^u Y^0
\left(\! \begin{array}{c} t^0_R \\ T^0_R \end{array}
\!\right) H + \text{H.c.} \notag \\
& = &
- \frac{1}{v} \left(\! \begin{array}{cc} \bar t_L & \bar T_L \end{array} \!\right)
\mathcal{M}^u_\text{diag} \, U_R^u \, Y^0 (U_R^u)^\dagger 
\left(\! \begin{array}{c} t_R \\ T_R \end{array}
\!\right) H + \text{H.c.} \,.
\end{eqnarray}
Then, the sum $Y_{tt}+Y_{TT}$ is simply the trace of either the matrix $Y=U_L^u Y^0 (U_L^u)^\dagger$ (singlets and triplets) or $Y=U_R^u Y^0 (U_R^u)^\dagger$ (doublets), which obviously equals unity.

In the down sector, on the other hand, the $b$ quark is much lighter and there is essentially no cancellation in the $gg\to H$ ($H \to gg$) and $H \to \gamma \gamma$ amplitudes because $|A_{1/2}(M_H^2/4 m_b^2)| \simeq 10^{-2}$, much smaller than $A_{1/2}(M_H^2/4 m_B^2) \simeq 4/3$. A good approximation is then obtained by using the heavy-quark limit for the $B$ quark and neglecting the contribution of the $b$ quark. Then, we get the new physics contribution to the amplitudes
\begin{equation}
F_d-F_d^\mathrm{SM}  \simeq \frac{4}{3} (s^d_{L,R})^2, 
\end{equation}
which is suppressed by the square of the mixing angle, $s^d_L$ for singlets and triplets and $s^d_R$ for doublets. We have seen in the previous section that the largest mixing allowed by electroweak tests for the down-type quarks occurs in the case of the \by\ doublet. With the largest allowed mixing, $s^d_R=0.18$, an exact calculation gives an increase in the $gg\to H$ cross section and $H \to gg$ partial width of 6.4\% with respect to the SM. In $H \to \gamma \gamma$, where the $W$~boson loop contributes dominantly, the partial width decreases by 0.4\% with respect to the SM. But another effect to account for is the change in the coupling of the Higgs to bottom quarks, which modifies at the tree level the decay width $H \to b \bar{b}$~\cite{Kearney:2012zi}. With the maximal mixing $s^d_R=0.18$, this partial width is reduced by a 6.4\% with respect to its SM value. Since this decay mode gives the bulk of the total width of the Higgs, with $\text{Br}(H \to b \bar b) = 0.578$ for $M_H=125$ GeV~\cite{Denner:2011mq,Dittmaier:2012vm}, the branching ratios into other final states are enhanced by an extra 3.8\%. In particular, the combined effect in the $H \to gg$ branching ratio is an increase by 10\% with respect to the SM value.

These effects are well below the precision of the current measurements~\cite{ATLAS:2013sla,CMS:yva}, and are likely invisible at the LHC. But they would be visible at the ILC, where the expected precision in Higgs branching ratio measurements is at the few percent level~\cite{DBD}. In particular, deviations in the branching ratios for $H \to gg$ and $H \to b \bar b$ would be at the $2\sigma$ level, given the expected precision $\Delta \text{Br}/\text{Br}$ of 4.8\% and 2.6\%, respectively, in their measurement.
Finally, note that different conclusions can be drawn in the presence of several different types of vector-like quarks with Yukawa couplings connecting them~\cite{Bonne:2012im,Almeida:2012bq,Batell:2012ca,Moreau:2012da} or in theories with large non-renormalisable couplings~\cite{Fajfer:2013wca}. Modifications of $H \to \gamma \gamma$ without affecting $gg \to H$, $H \to gg$ are possible with extra vector-like leptons~\cite{Ishiwata:2011hr,Carena:2012xa,Joglekar:2012vc,ArkaniHamed:2012kq}.


\section{Heavy quark decays}
\label{sec:split}
\label{sec:dec}

\begin{figure}[htb]
\begin{center}
\begin{tabular}{cc}
\includegraphics[height=5.4cm,clip=]{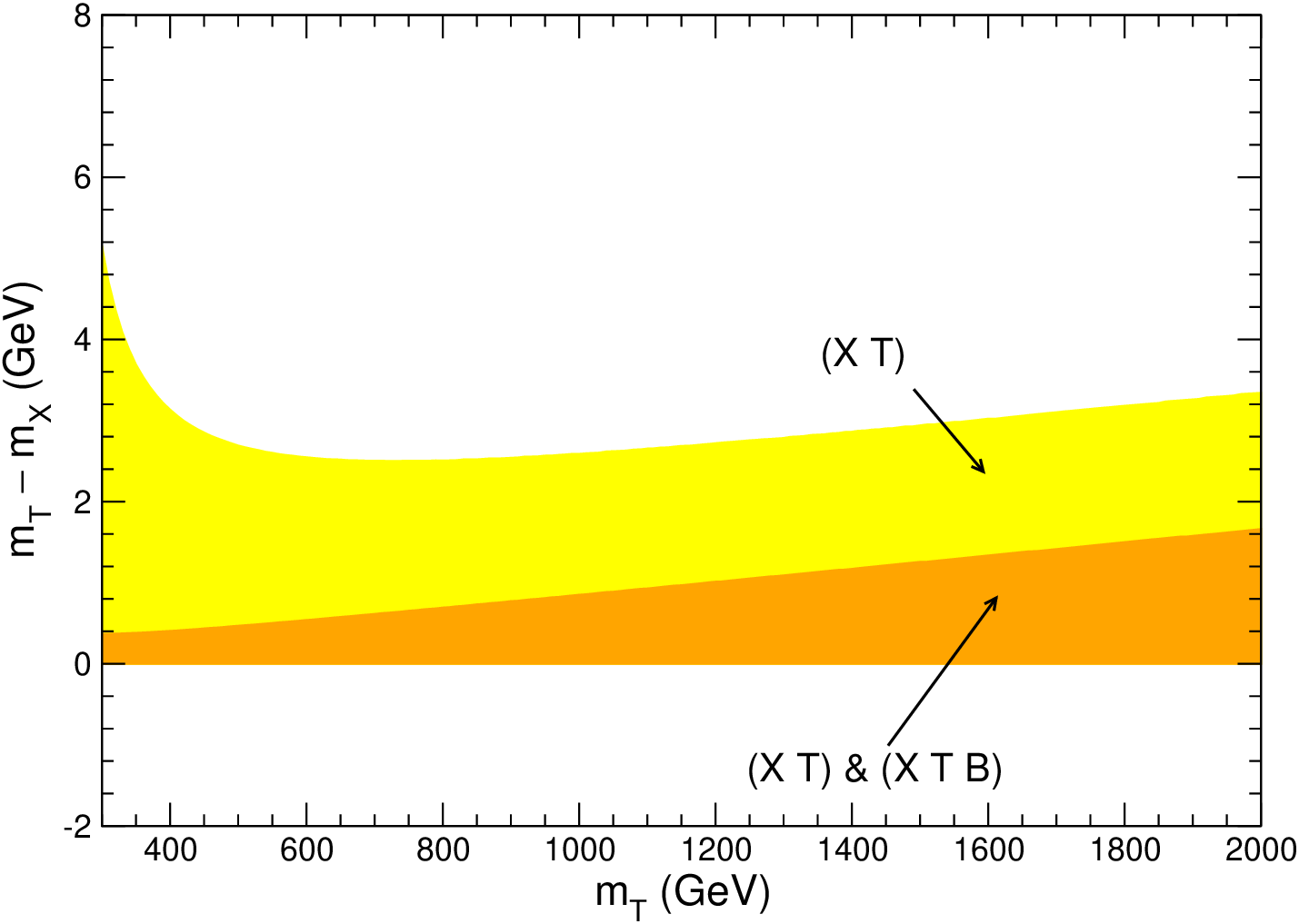} & \includegraphics[height=5.4cm,clip=]{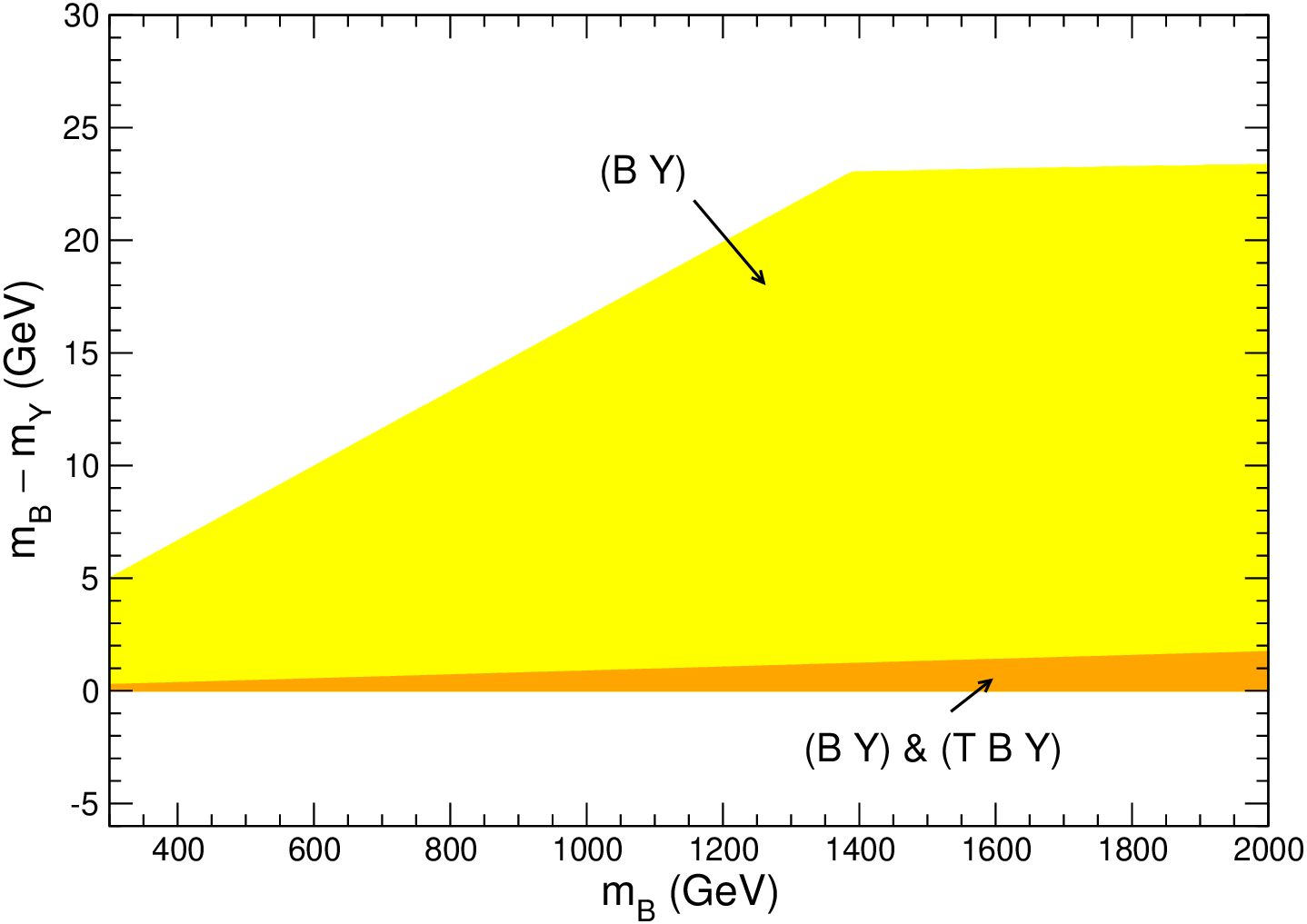} \\
\multicolumn{2}{c}{\includegraphics[height=5.4cm,clip=]{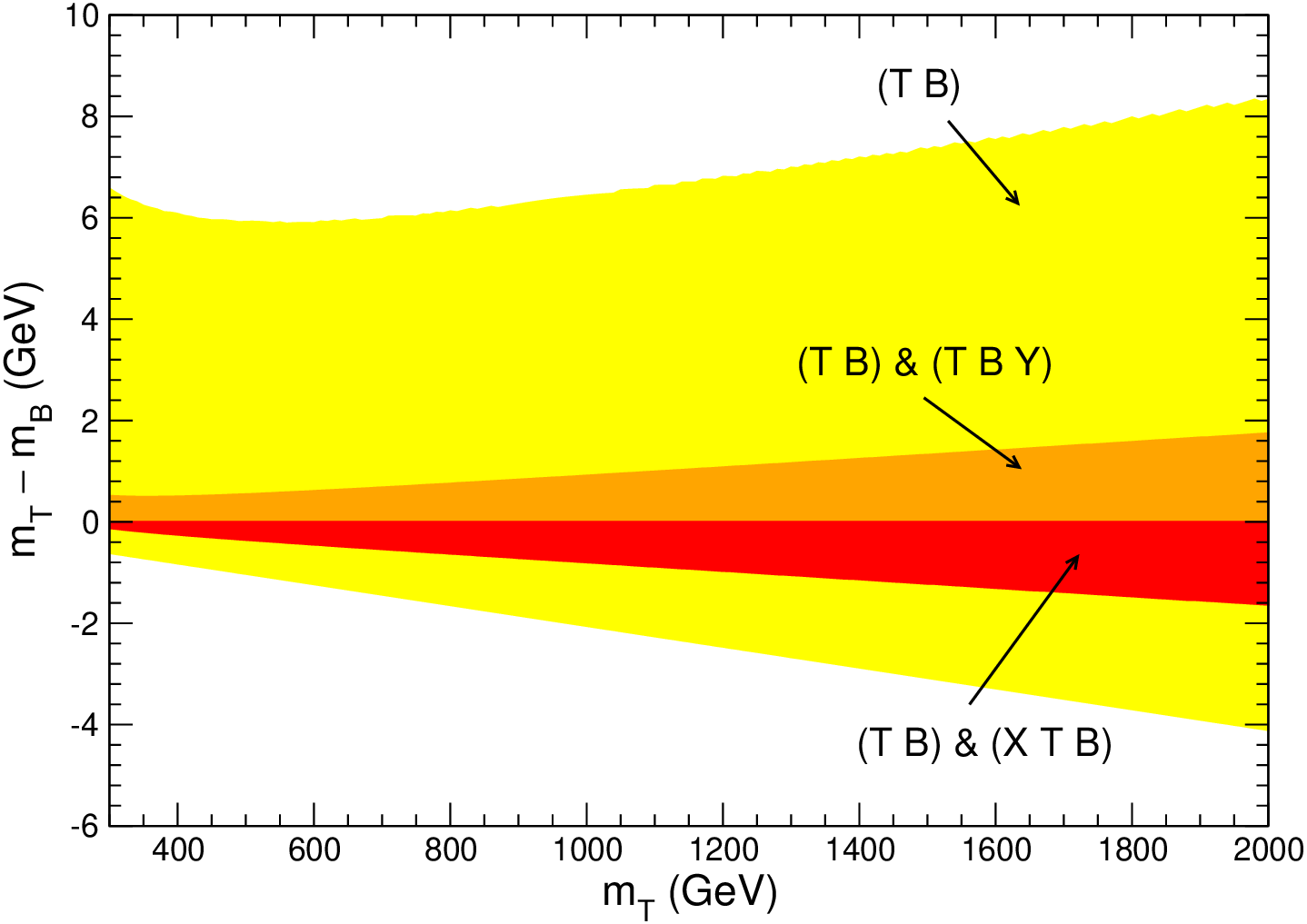}}
\end{tabular}
\caption{Allowed range for the splitting of the heavy quark masses.}
\label{fig:split}
\end{center}
\end{figure}

In multiplets with more than one heavy particle (doublets, triplets) the two heavy states share the same bare mass term, but the mixing with lighter partners $t,b$ induces a splitting of their mass, given by the equations in Table~\ref{tab:split}.
\begin{table}[htb]
\begin{center}
\begin{tabular}{cl}
\xt & $m_X^2 = m_T^2 \crx^2 + m_t^2 \srx^2$ \\
\tb & $m_T^2 (\cru)^2 + m_t^2 (\sru)^2 = m_B^2 (\crd)^2 + m_b^2 (\srd)^2$ \\
\by & $m_Y^2 = m_B^2 \crx^2 + m_b^2 \srx^2$ \\
\xtb & $m_X^2 = m_T^2 (\clu)^2 + m_t^2 (\slu)^2$ \\
       & $m_T^2 (\clu)^2 + m_t^2 (\slu)^2 = m_B^2 (\cld)^2 + m_b^2 (\sld)^2$ \\
\tby & $m_Y^2 = m_B^2 (\cld)^2 + m_b^2 (\sld)^2$ \\
       & $m_T^2 (\clu)^2 + m_t^2 (\slu)^2 = m_B^2 (\cld)^2 + m_b^2 (\sld)^2$
\end{tabular}
\caption{Splitting between heavy quark masses for the doublets and triplets.}
\label{tab:split}
\end{center}
\end{table}
In particular, one can easily observe from these equations that $m_T \geq m_X$, $m_B \geq m_Y$ while $T$ can be heavier or lighter than $B$. The allowed range of the splitting, given the constraints on mixing in Section~\ref{sec:lim}, are presented in Fig.~\ref{fig:split}. Notice that in the three plots the allowed regions for the triplets are fully contained in the ones for the doublets, as indicated by the labels. Moreover, in the lower plot the regions for the two triplets are disjoint. 

The small mass difference between the heavy members of the multiplets suppresses the decay from one to the other. Hence, the only possible decays for the heavy states are into top / bottom quarks plus a $W$, $Z$ or Higgs boson. For the quarks with exotic charges $5/3$, $-4/3$ the only decay channels are $X \to W^+ t$, $Y \to W^- b$, with total widths given in Appendix~\ref{sec:b}. For the heavy quarks with charges $2/3$ and $-1/3$, the possible channels are well
known~\cite{delAguila:1989rq},
\begin{align}
& T \to W^+ b \,,\quad T \to Zt \,,\quad T \to Ht \,, \notag \\
& B \to W^- t \,,\quad B \to Zb \,,\quad B \to Hb \,.
\end{align}
The partial widths for all these modes are also collected in Appendix~\ref{sec:b}.
The branching ratios for the different channels have some dependence on the heavy quark masses, resulting from kinematics. In all multiplets except the \tb\ doublet, there is only one independent mixing parameter, and the dependence of the branching ratios on its value is marginal because one of the chiralities is always very suppressed with respect to the other one and, given the constraints presented in Section \ref{sec:lim}, the dominant charged current and neutral mixings are similar, $X \simeq V$, as it can be checked with the explicit expressions given in Appendix~\ref{sec:a}. For the \tb\ doublet the branching ratios of $T,B$ do depend on the relative size of $\theta_R^u$ and $\theta_R^d$. We thus have considered three scenarios: (i) $\theta_R^u \neq 0$, $\theta_R^d=0$ (labelled as `d0'), $\theta_R^d \neq 0$, $\theta_R^u = 0$ (`u0') and maximal mixing (maximum values of $\sin \theta_R^u$ and $\sin \theta_R^d$ in Fig.~\ref{fig:lim}, for a given $m_T$), labelled as `max'.

\begin{figure}[htb]
\begin{center}
\begin{tabular}{cc}
\includegraphics[height=7.5cm,clip=]{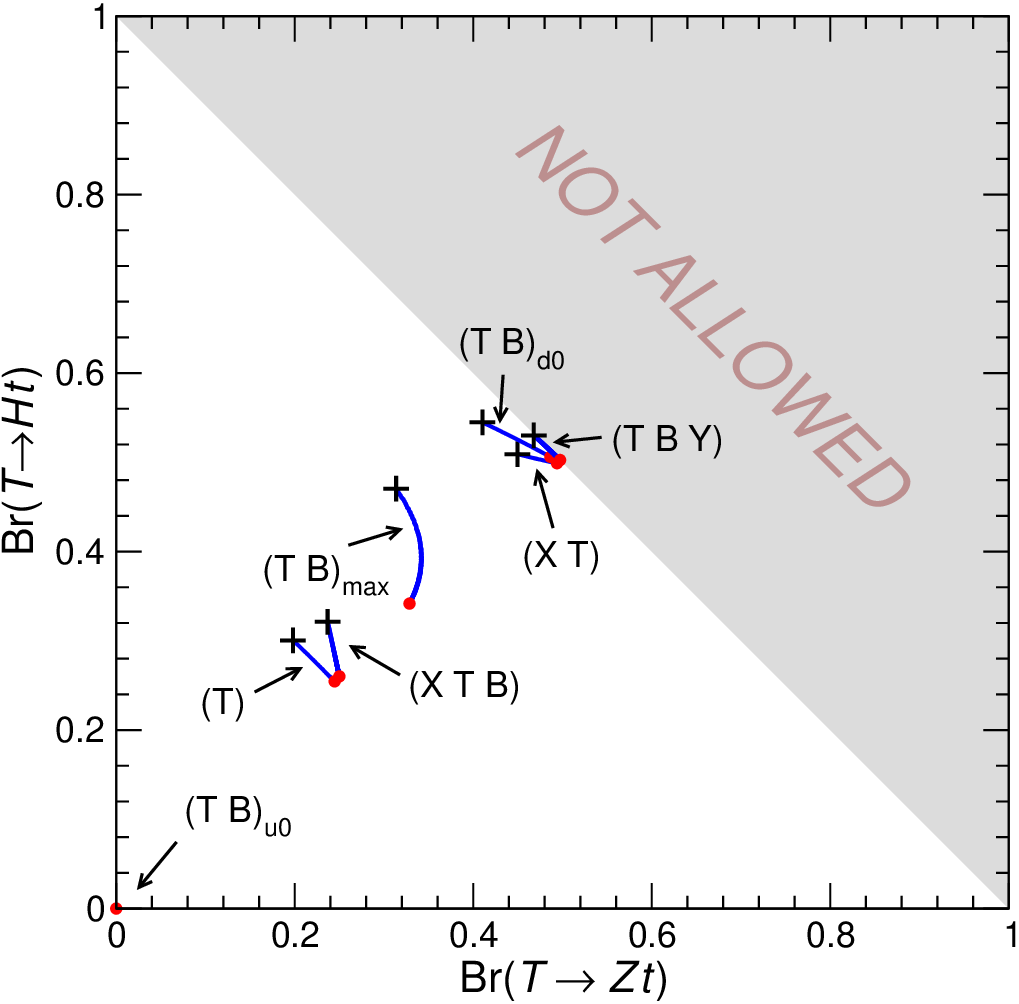} & \includegraphics[height=7.5cm,clip=]{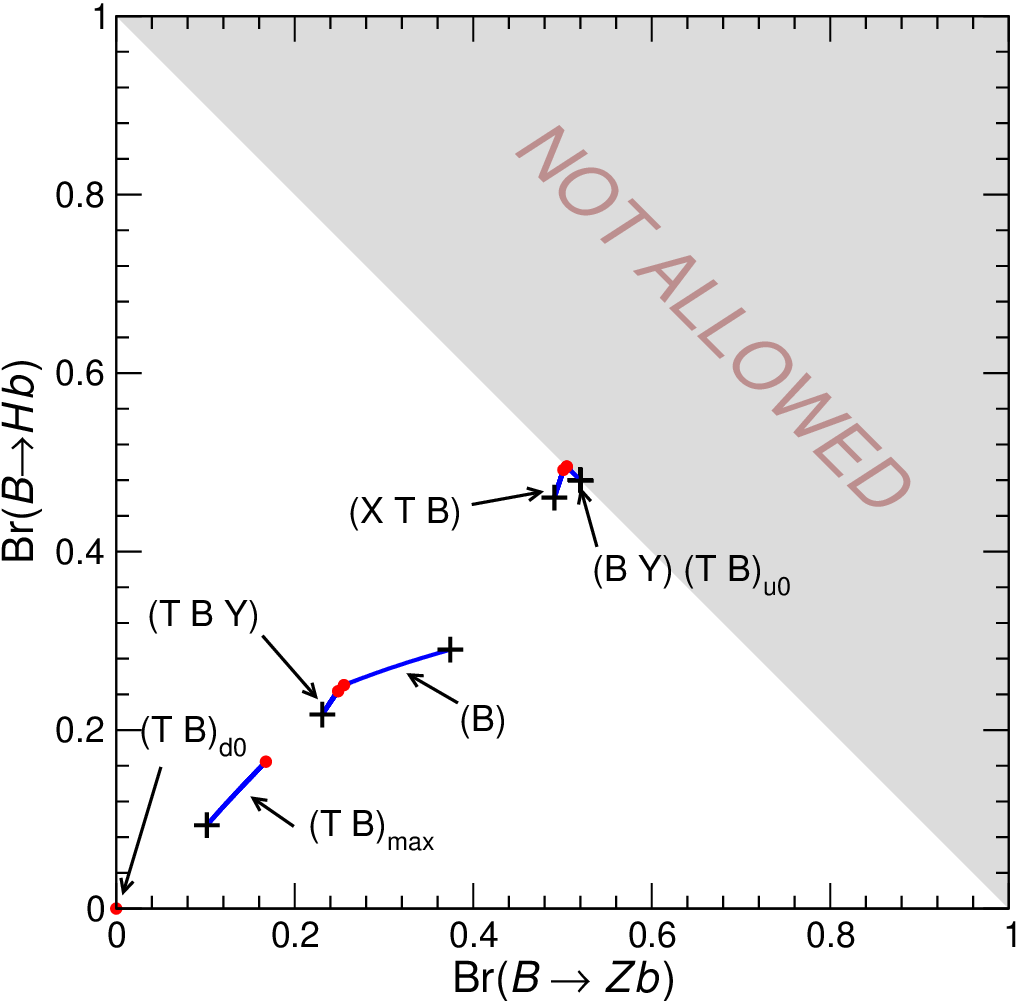} 
\end{tabular}
\caption{Allowed branching ratios for the decays of $T$ (left) and $B$ (right) quarks in the different multiplets.}
\label{fig:BR}
\end{center}
\end{figure}

The allowed branching ratios for $T$, $B$ in the different multiplets are presented in Fig.~\ref{fig:BR}. 
The three branching ratios are not independent, since
\begin{equation}
\text{Br}(Q \to Wq') + \text{Br}(Q \to Zq) + \text{Br}(Q \to Hq) =1 \,,\quad (Q,q,q')=(T,t,b),(B,b,t) \,.
\label{ec:sumBR}
\end{equation}
The values of $\text{Br}(Q \to Zq)$ and $\text{Br}(Q \to Hq)$, are given in the horizontal and vertical axes, respectively, while the charged current one is obtained by the condition in Eq.~(\ref{ec:sumBR}). The dots represent the values for $m_Q = 2$ TeV, and are very close to the asymptotic values for very heavy quark masses where $\lambda \simeq m_Q^2$, $r_x \simeq 0$. The crosses indicate the points corresponding to the current mass limit, which depend on the specific decay channels. Currently, the limits are
\begin{align}
& m_X > 770~\text{GeV \cite{CMS:vwa}}  && (X\,T)\,,~(X\,T\,B) \,, \notag \\
& m_T > 640~\text{GeV \cite{ATLAS:2013ima}} && T\,,~(X\,T\,B) \,, \notag \\
& m_T > 790~\text{GeV \cite{ATLAS:2013ima}} && (X\,T)\,,~(T\,B\,Y) \,, \notag \\
& m_T > 640~\text{GeV \cite{ATLAS:2013ima,ATLAS:2012qe}} && (T\,B) \,, \notag \\
& m_B > 590~\text{GeV \cite{ATLAS:bp}} && B\,,~(T\,B\,Y) \,, \notag \\
& m_B > 358~\text{GeV \cite{Aad:2012pga}} && (T\,B)\,,~(B\,Y) \,, \notag \\
& m_Y > 656~\text{GeV \cite{ATLAS:2012qe}} && (B\,Y)\,,~(T\,B\,Y) \,.
\end{align}
The lines between dots and crosses represent the branching ratios for intermediate masses. Note that for a \tb\ doublet with $\sin \theta_R^u=0$ the $T$ quark does not couple to $Z$ and $H$, hence it only decays to $W^+b$. Conversely, for  $\sin \theta_R^d=0$ the $B$ quark does not couple to $Z,H$ and can only decay into $W^-t$.
We also remark that for the $T$ and $B$ quarks appearing in the two triplet representations the decay branching ratios are very similar to the ones for singlets or doublets. Therefore, the study of heavy quark pair production in~\cite{AguilarSaavedra:2009es} can be trivially extended to these cases too. Also, in Fig.~\ref{fig:BR}
it can be noticed that in both triangles the allowed branching ratios lie around the lines between $(0,0)$ and $(0.5,0.5)$. This is a consequence of the exact equality between the moduli of $Y_{qQ}$ and the dominant $X_{qQ}$ coupling (see Appendix~\ref{sec:a}).


\section{Single production at LHC}
\label{sec:xsec}

One of the most interesting consequences of our limits concerns the single production of heavy quarks at the LHC. The cross sections for these processes are proportional to the square of the couplings to the $W$ or $Z$ bosons, hence the limits obtained in Section~\ref{sec:lim} determine the maximum cross section for these processes.

Cross sections for the different processes of single vector-like quark production have been previously obtained in~\cite{AguilarSaavedra:2009es}.
Heavy quarks $T,\,Y$ that couple to the $b$ quark and the $W$ boson can be produced in $T \bar b j$, $Y \bar b j$, being $j$ a light quark jet, with relatively large cross sections. (The charge conjugate processes are always understood, and their cross sections included in the results presented.) These processes take place via the exchange of a $t$-channel $W$ boson, in full analogy with $t$-channel single top production in the SM. $B$ quarks can be produced in a similar process, $B \bar b j$, with the exchange of a $t$-channel $Z$ boson, but with lower cross sections for equal mixings. Charge $5/3$ quarks can only be produced in association with a top quark, $X \bar t j$, involving a $t$-channel $W$ boson, but the cross section is much smaller than for the former processes. Even smaller is the cross section for $T \bar t j$ with $Z$ boson exchange, which is the only production process for a $T$ quark with very small coupling to the $W$, as for example in the case of the \xt\ doublet.

\begin{figure}[htb]
\begin{center}
\begin{tabular}{cc}
\includegraphics[height=5.4cm,clip=]{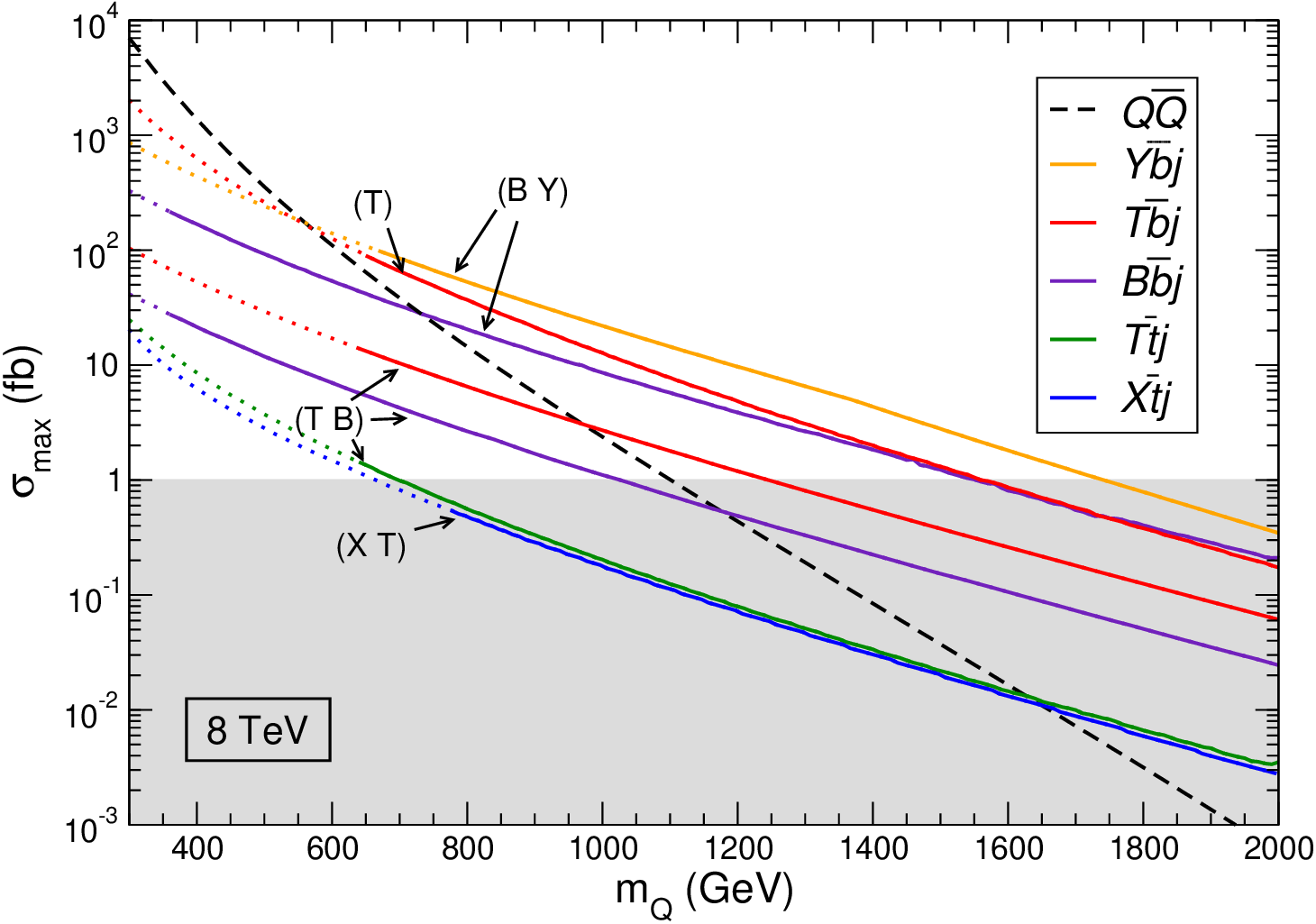} & \includegraphics[height=5.4cm,clip=]{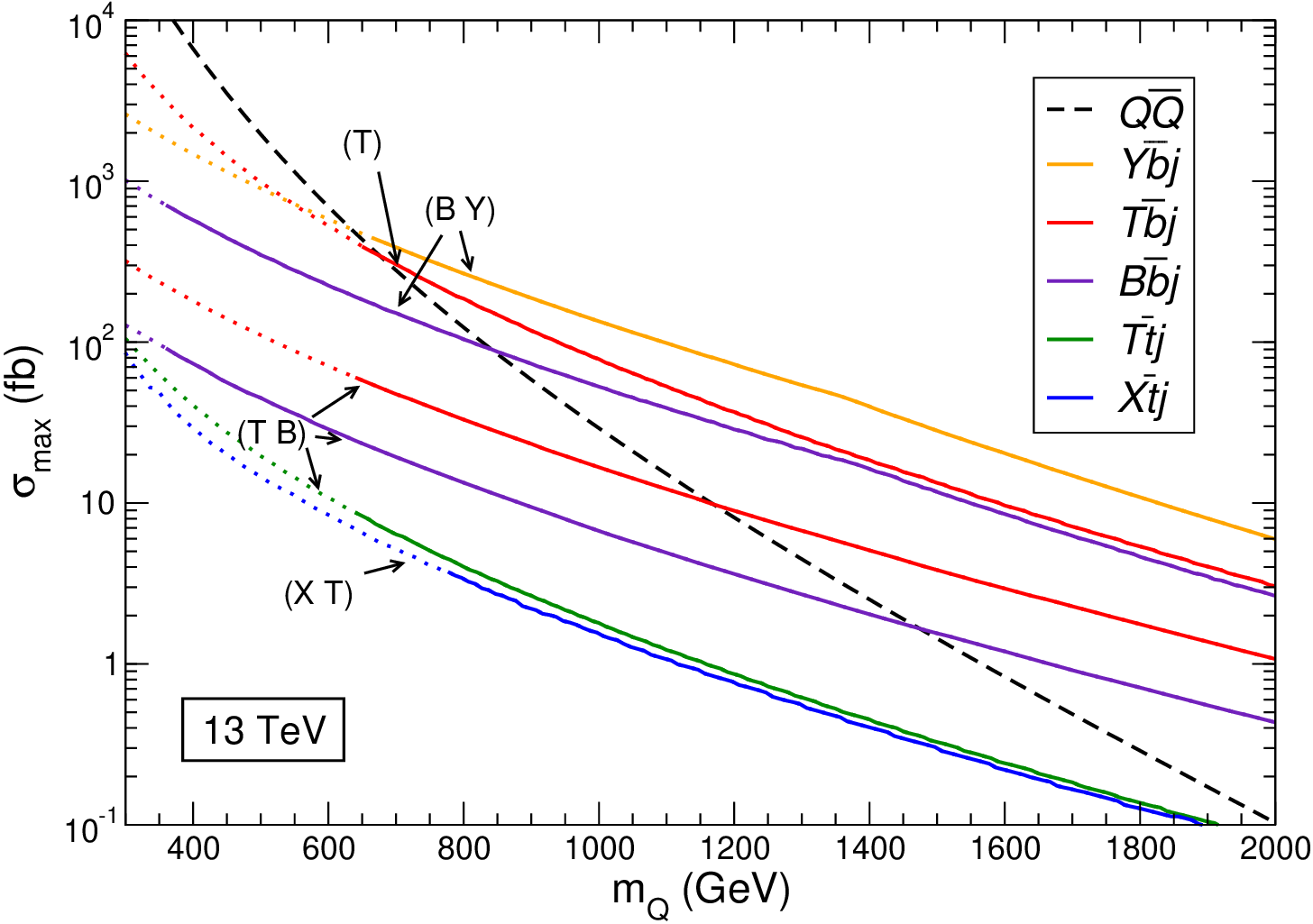}
\end{tabular}
\caption{Maximum single heavy quark production cross sections at the LHC with 8 TeV (left) and 13 TeV (right), for selected quark multiplets. The dotted part of the lines indicate the range of masses already excluded by direct searches. In the left plot, the shaded area corresponds to cross sections below 1 fb, uninteresting for the luminosity $L \simeq 20$ fb$^{-1}$ collected. }
\label{fig:xsec}
\end{center}
\end{figure}

The maximum cross sections for the most interesting processes (corresponding to the multiplets with largest mixing) are presented in Fig.~\ref{fig:xsec} for centre of mass (CM) energies of 8 TeV (left) and 13 TeV (right). They have been computed with {\tt Protos}~\cite{AguilarSaavedra:2009es} at the tree level. (Next-to-leading order calculations for single $T$ production are available~\cite{Berger:2009qy,Campbell:2009gj}.) They comprise:
\begin{itemize}
\item $T \bar b j$ for the $T$ singlet
\item $T \bar b j$, $B \bar b j$ and $T \bar t j$ for the \tb\ doublet. Remarkably, the cross section for $T \bar b j$, which is proportional to the mixing in the {\it down} sector, can be larger than the one for $T \bar t j$, which is proportional to the mixing in the up sector but is a much more suppressed process. $B \bar b j$ is also proportional to the mixing in the down sector, but its cross section is smaller, as mentioned above.
\item $Y \bar b j$ and $B \bar b j$ for the \by\ doublet.
\item $X \bar t j$ for the \xt\ doublet. The cross section for $T \bar t j$ in this model is even smaller.
\end{itemize}
In the plots, for a given value of the mass the mixing is set to the maximum allowed by indirect constraints in Fig.~\ref{fig:lim}, thereby obtaining the maximum cross section for each process. For comparison we also include the pair production cross section, which is independent of the mixing. Single production for the rest of multiplets is small, and can be estimated from the data shown in Fig.~\ref{fig:xsec} and the limits in Section~\ref{sec:lim}.

These results deserve a detailed discussion. We observe that $Y \bar b j$, $T \bar b j$ and $B \bar b j$ are the only relevant single-production processes at 8 TeV, since the rest have cross sections that are below the one for pair production. The same can be said for a CM energy of 13~TeV. Precisely these three processes involve Feynman diagrams with initial state gluon splitting $g \to b \bar b$, as $t$-channel single top production in the SM.
 The $X \bar t j$ and $T \bar t j$ processes, involving initial gluon splitting $g \to t \bar t$, always have cross sections far below the one for pair production in minimal models with only one multiplet.\footnote{These processes might be enhanced in models that evade the limits in Section~\ref{sec:lim} via cancellations of the contributions of different vector-like multiplets (and/or other types of new physics), as we have discussed above. Whether the large mixings necessary to make these processes phenomenologically relevant are compatible with precision electroweak data needs to be checked for each model of this kind.}

The $Y$ quark decays into $W^- b$ with 100\% branching ratio, so the signal resulting from its single production is $Y \bar b j \to W^- b \bar b j$, which may be distinguished from the production of $W+\text{jets}$ by the large $Wb$ invariant mass and the presence of a forward jet. For $T \bar b j$, the $T$ singlet decays into $W^+ b$, $Zt$ and $Ht$ with branching ratios around $0.5:0.25:0.25$. The resulting signal $W^+b \bar b j$ should be visible over the $W+$ jets background; in the $Zt$ decay channel the leptonic $Z$ mode gives a clean signal but with a small branching ratio and the signal in the Higgs channel might be identified by requiring several $b$ tags and a forward jet. The same can be said about $B \bar b j$ with $B \to Hb,Zb$, which have branching ratios around $0.5:0.5$ for the \by\ doublet. More detailed studies of the LHC sensitivity to single $T$ production have been given in~\cite{Azuelos:2004dm,Yue:2009cq,Vignaroli:2012sf}.


\section{Effects in top couplings}
\label{sec:top}

\begin{figure}[htb]
\begin{center}
\begin{tabular}{cc}
\includegraphics[height=5.2cm,clip=]{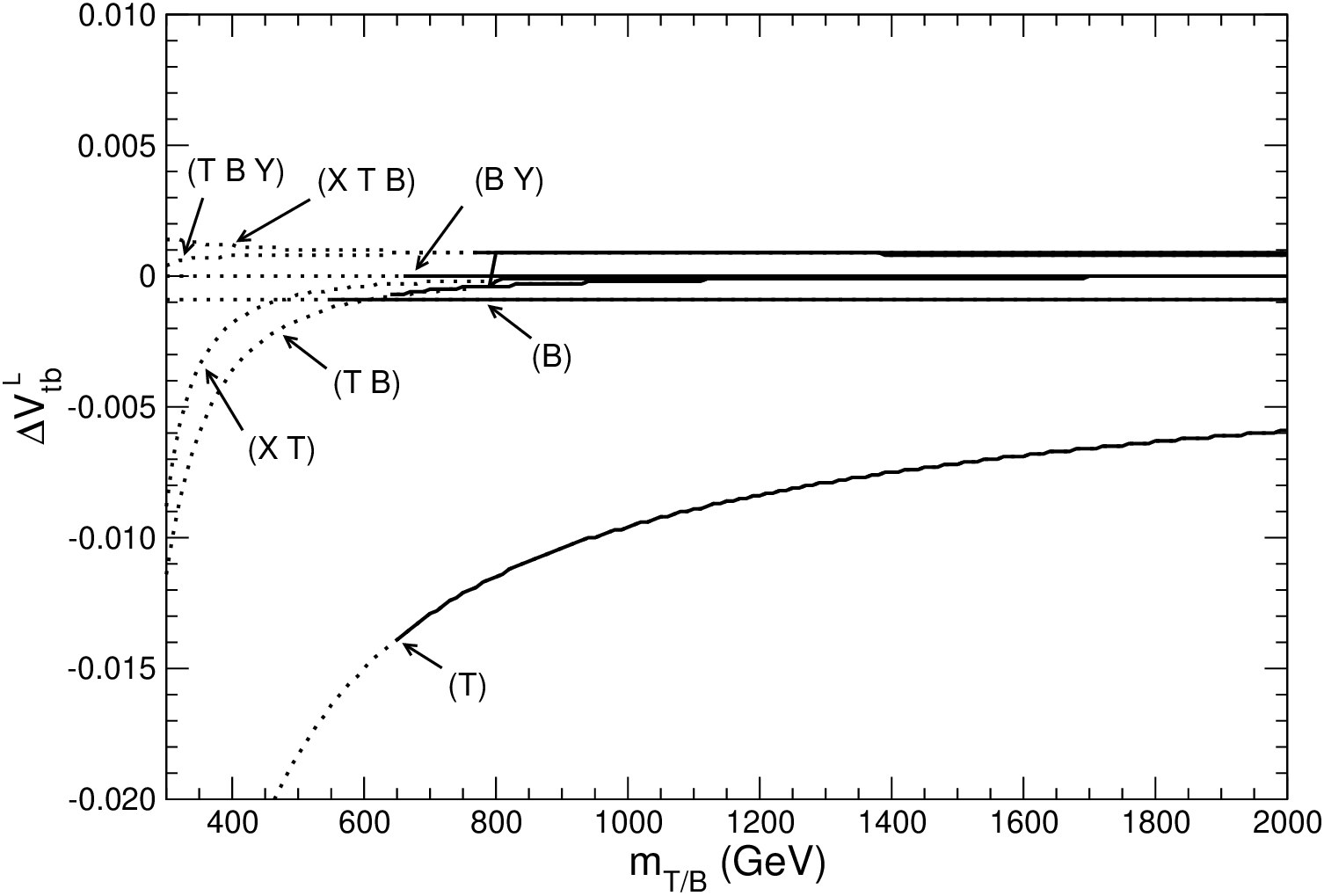} & \includegraphics[height=5.2cm,clip=]{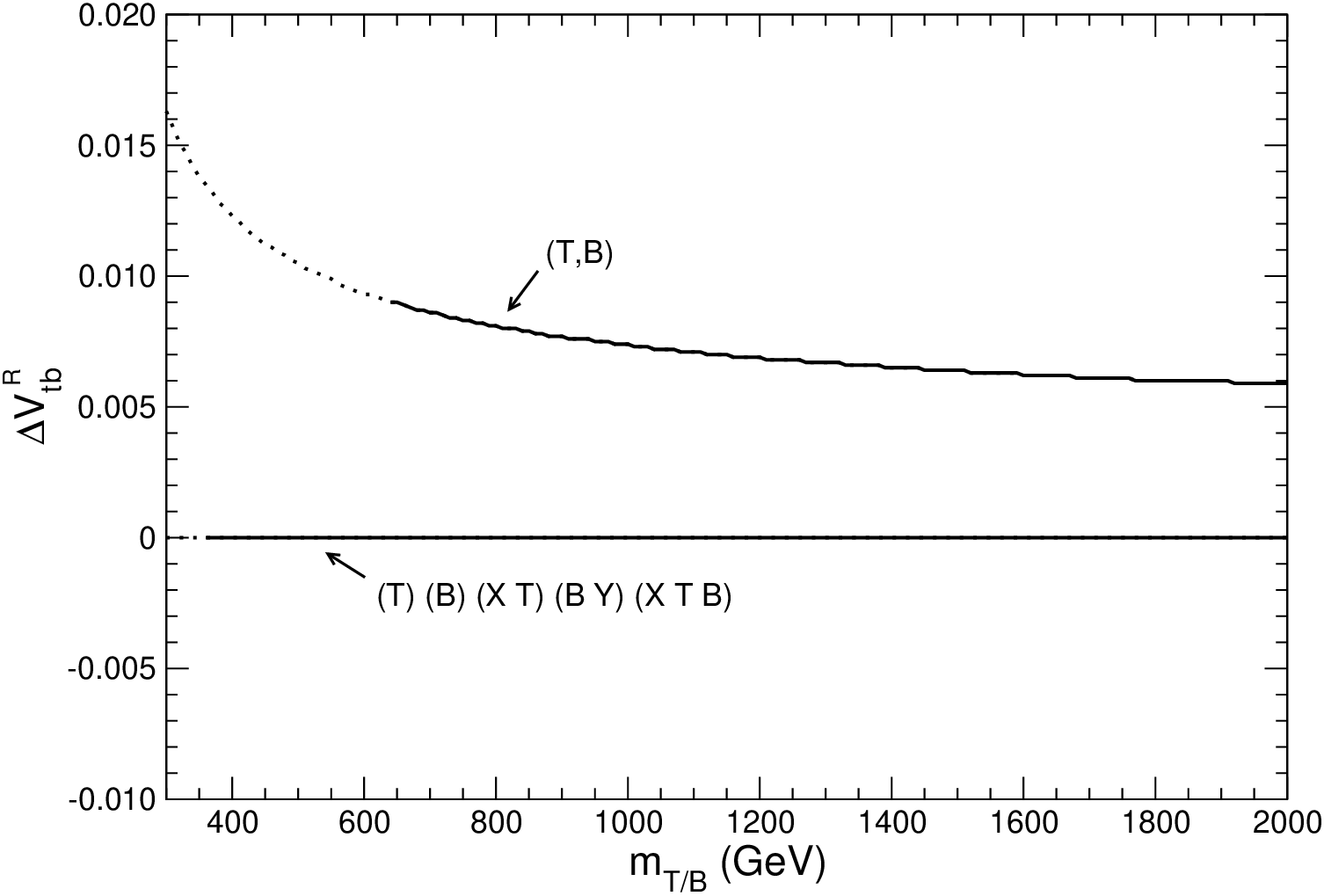} \\
\includegraphics[height=5.2cm,clip=]{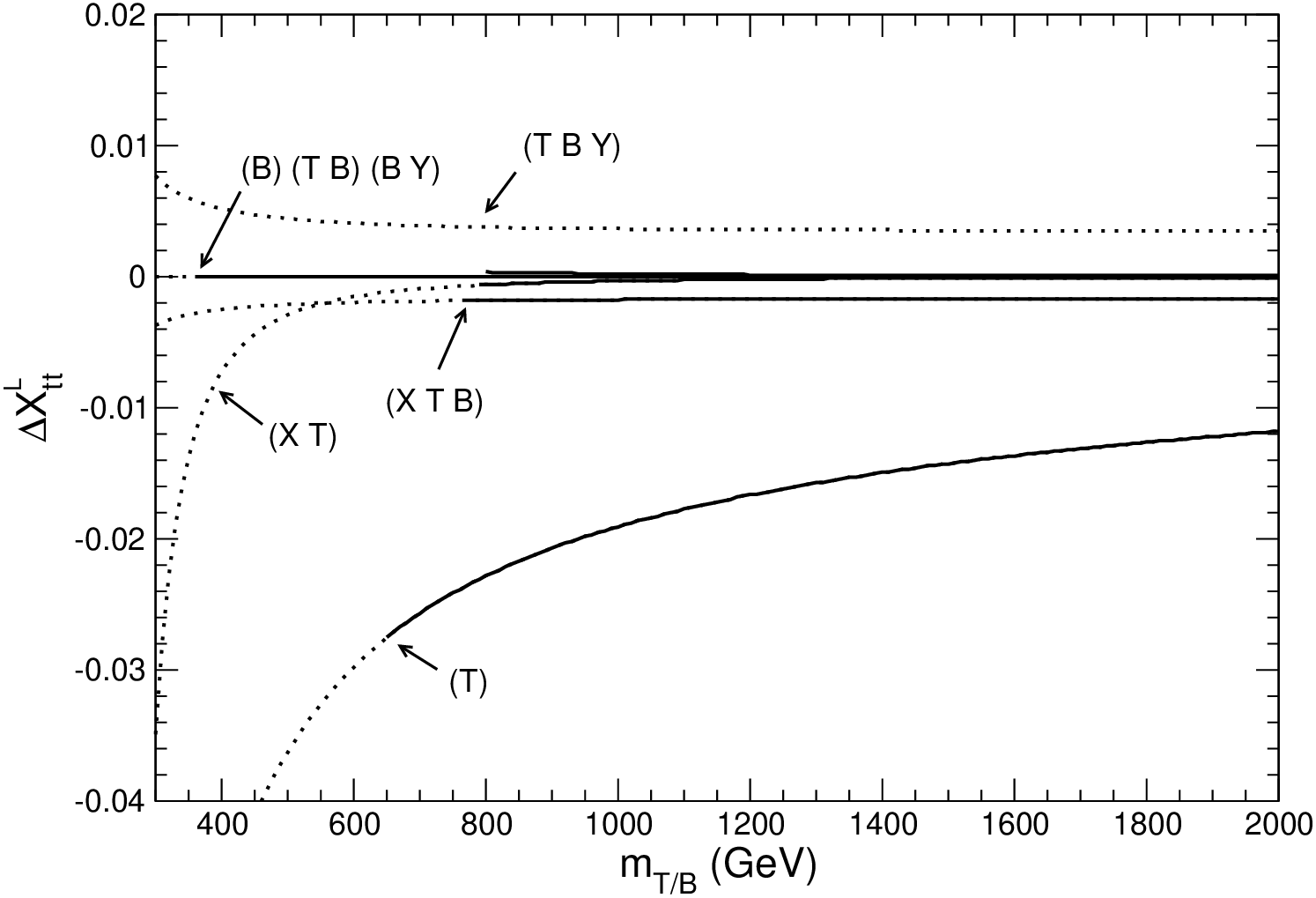} & \includegraphics[height=5.2cm,clip=]{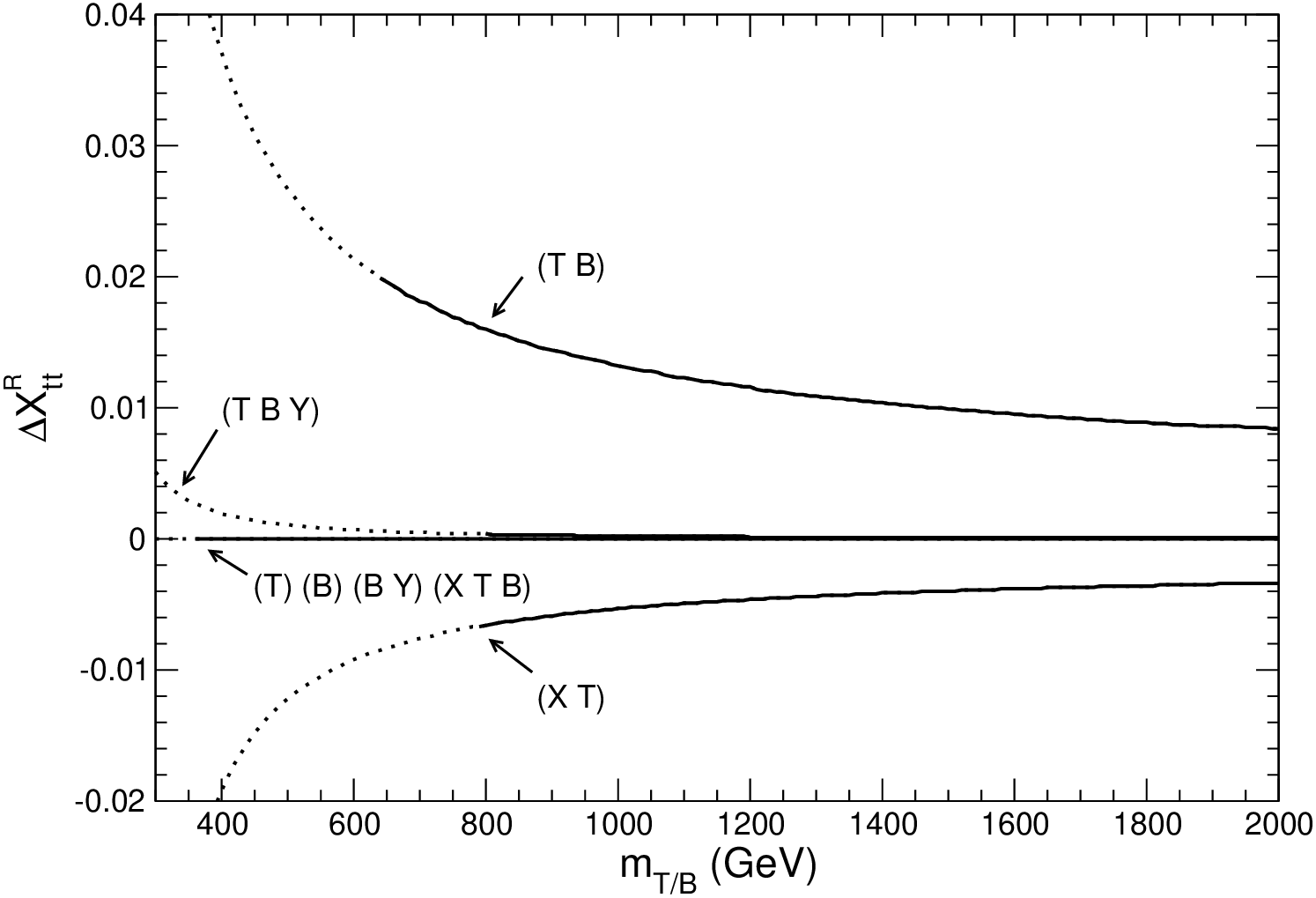} \\
\multicolumn{2}{c}{\includegraphics[height=5.2cm,clip=]{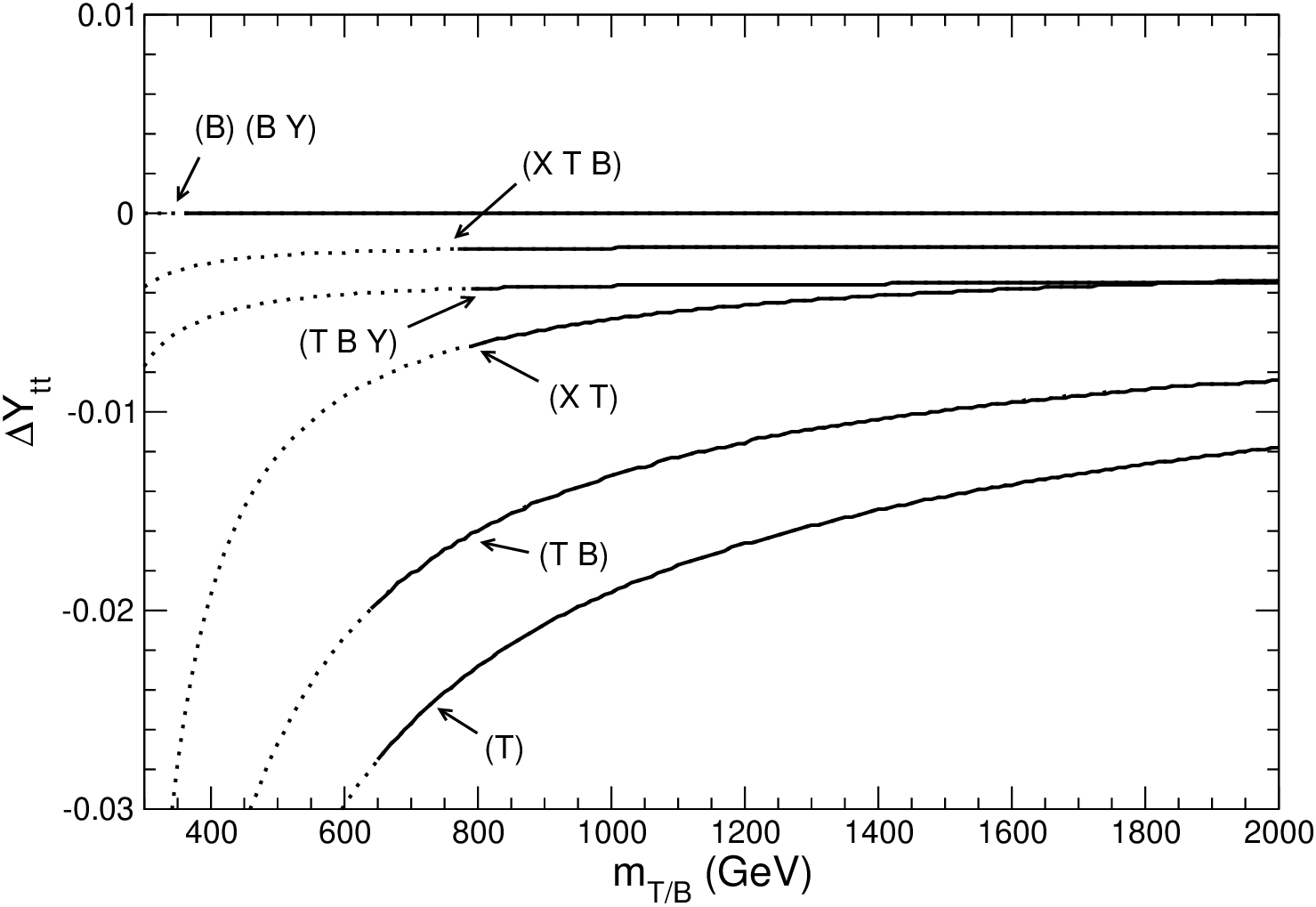}}
\end{tabular}
\caption{Allowed deviations in top couplings. The dotted part of the lines indicates the range of masses already excluded by direct searches.}
\label{fig:mix}
\end{center}
\end{figure}

Using the explicit expressions for the Lagrangians collected in Appendix~\ref{sec:a}, 
the limits on mixing angles presented in the previous section can be directly translated into constraints on the possible deviations of the top couplings to the $W$, $Z$ and Higgs bosons,
\begin{align}
& \Delta V_{tb}^L \equiv V_{tb}^L - (V_{tb}^L)_\text{SM} \simeq V_{tb}^L -1 \,,
&& \Delta V_{tb}^R \equiv V_{tb}^R - (V_{tb}^R)_\text{SM} = V_{tb}^R \,, \notag \\
& \Delta X_{tt}^L \equiv X_{tt}^L - (X_{tt}^L)_\text{SM} = X_{tt}^L -1 \,,
&& \Delta X_{tt}^R \equiv X_{tt}^R - (X_{tt}^R)_\text{SM} = X_{tt}^R \,, \notag \\
& \Delta Y_{tt} \equiv Y_{tt} - (Y_{tt})_\text{SM} \,.
\end{align}
The results are presented in Fig.~\ref{fig:mix}.
The deviations in $V_{tb}^L$ are too small to be observed in single top production at the LHC, given the present size of systematic uncertainties, around 7\% in the best case~\cite{Chatrchyan:2012ep,Aad:2012ux}. 
Likewise, the possible appearance of a right-handed coupling $V_{tb}^R$ would not show up in measurements of $W$ helicity fractions and related observables given the current sensitivity to $\Delta V_{tb}^R \sim 0.2$~\cite{Aad:2012ky} or even with the envisaged precision $\Delta V_{tb}^R \sim 0.06$~\cite{AguilarSaavedra:2007rs}.
The deviations in the Higgs Yukawa coupling of the top are also very small, well below the expected precision at the LHC, $\Delta Y_{tt}\sim 0.2$~\cite{AllwoodSpiers:2012zz} and even at the ILC, $\Delta Y_{tt}\sim 0.1$~\cite{Martinez:2002st,Yonamine:2011jg}.

\begin{figure}[htb]
\begin{center}
\begin{tabular}{cc}
\includegraphics[height=5.3cm,clip=]{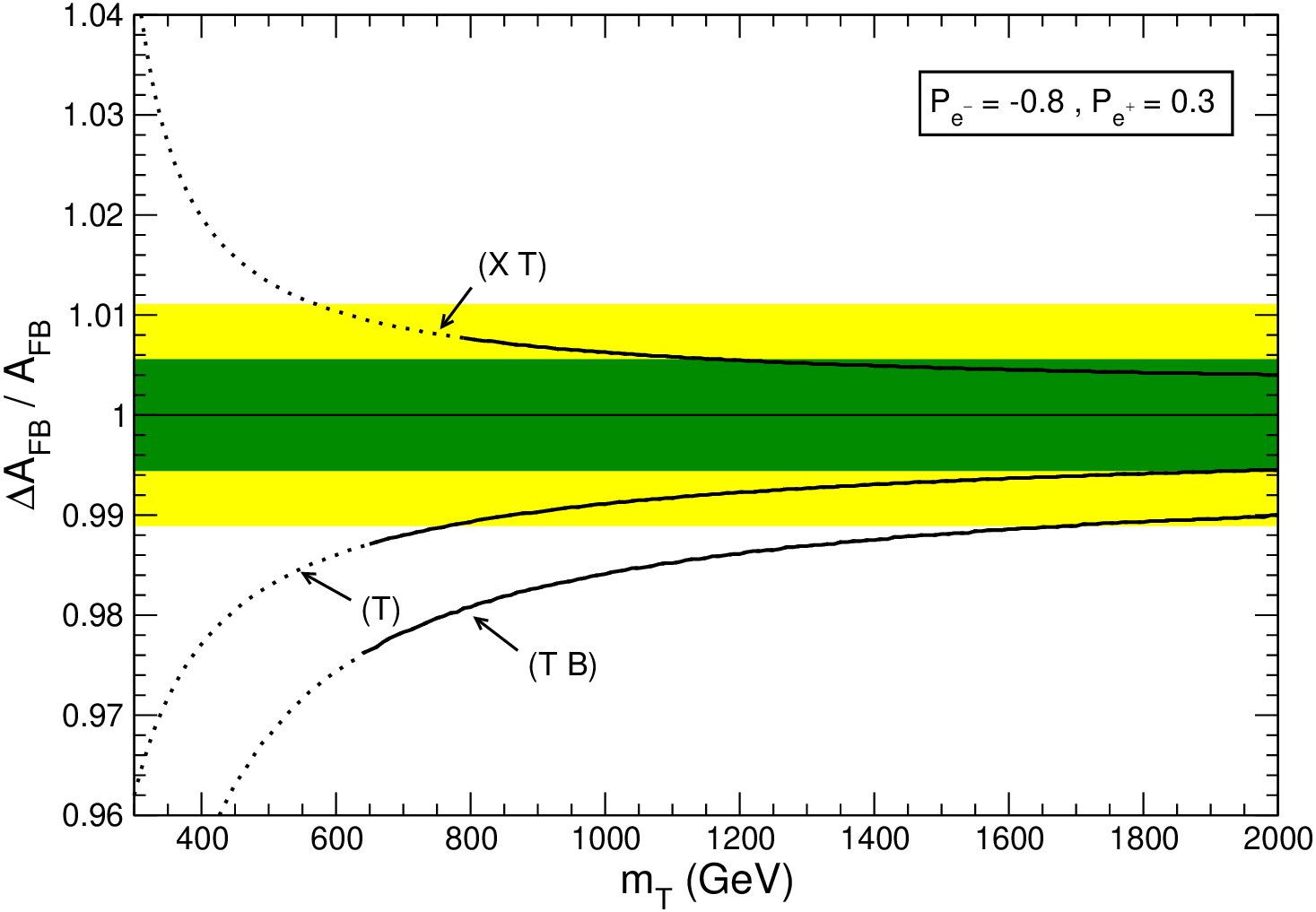} & \includegraphics[height=5.3cm,clip=]{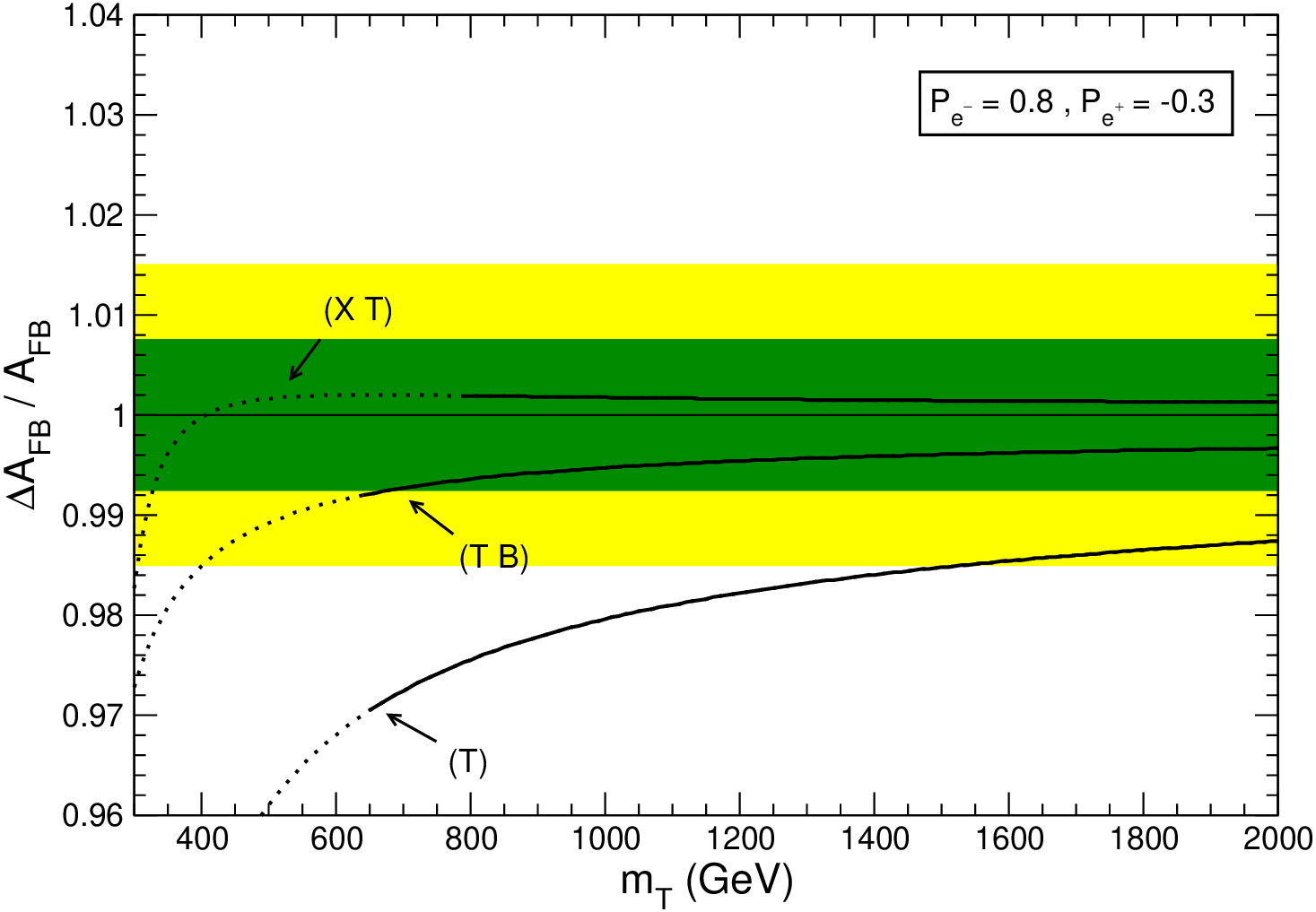} \\
\multicolumn{2}{c}{\includegraphics[height=5.3cm,clip=]{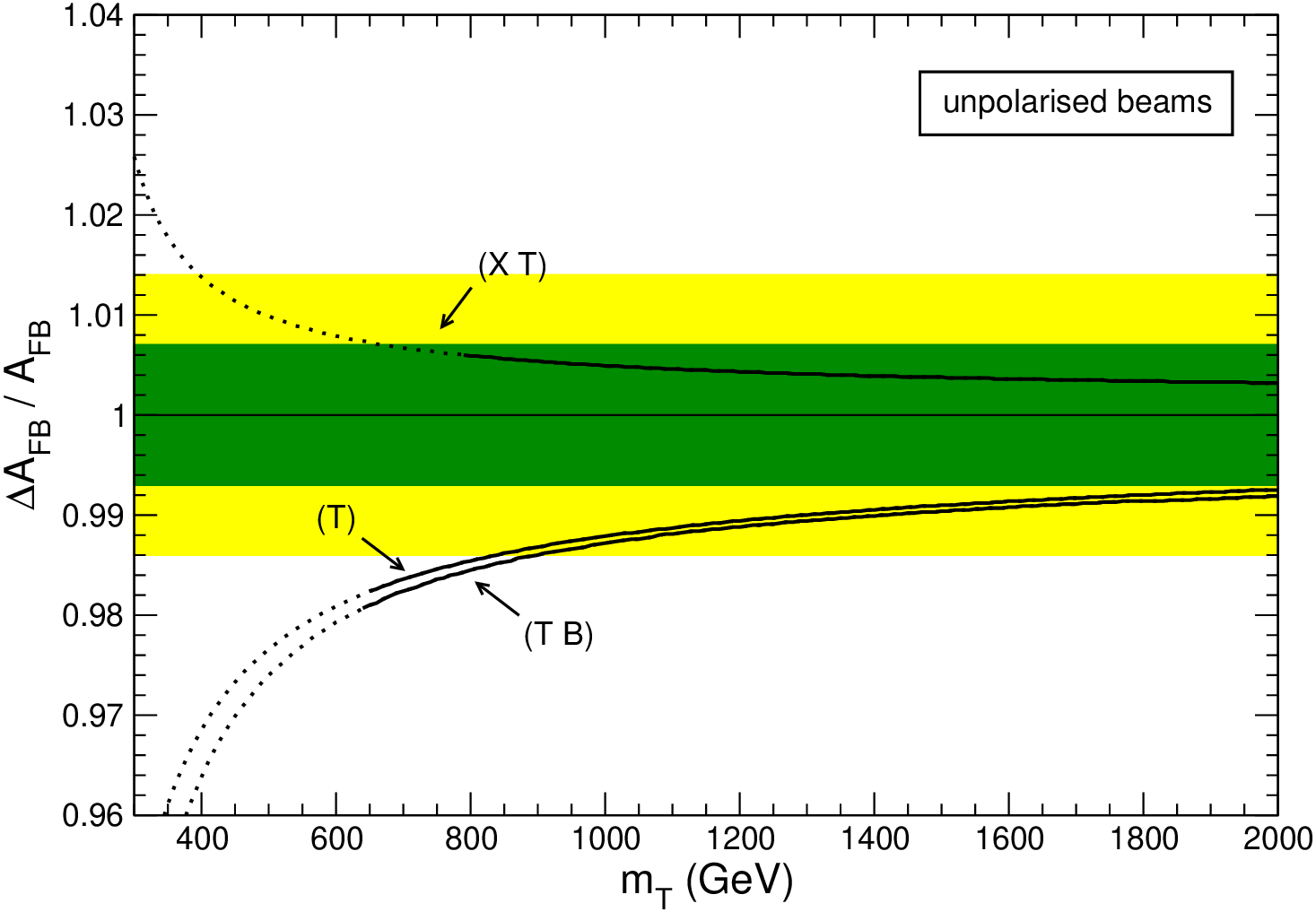}}
\end{tabular}
\caption{Deviations in the FB asymmetry in $e^+ e^- \to t \bar t$ at the ILC. The dotted part of the lines indicates the range of masses already excluded by direct searches.}
\label{fig:ILC}
\end{center}
\end{figure}

On the other hand, the couplings to the $Z$ boson are expected to be measured with a very good precision at the ILC. As an example, we show in Fig.~\ref{fig:ILC} the variation of the FB asymmetry in $e^+ e^- \to t \bar t$ (see~\cite{AguilarSaavedra:2012vh}) for three polarisation options: (i) $P_{e^-} = -0.8$, $P_{e^+} = 0.3$;  (ii) $P_{e^-} = 0.8$, $P_{e^+} = -0.3$; (iii) no polarisation.
The green and yellow bands represent the $1\sigma$ and $2\sigma$ statistical uncertainty, taking a luminosity of 500~fb$^{-1}$ and a bulk detection efficiency of 25\% in the semileptonic $t \bar t$ decay channel, which is similar to the one achieved at the LHC~\cite{Aad:2012qf,Chatrchyan:2012ria}. Systematic uncertainties are not included and would slightly degrade the sensitivity. We observe that detection of indirect effects of quark mixing could in principle be possible, even if the new quarks are beyond the LHC reach. But this would demand keeping systematic uncertainties in these asymmetries below 1\%, which requires a very good reconstruction of the $t \bar t$ pair~\cite{Doublet:2012wf}.

\section{Effects in bottom couplings: improving the electroweak fit}
\label{sec:fit}

In the bottom sector, there is already a deviation that demands an explanation: the FB asymmetry in $e^+ e^- \to Z \to b \bar b$ at LEP~\cite{ALEPH:2005ab}. Actually, the measured $R_b$ is above the SM prediction, while $A_\text{FB}^b$ is below it. Hence, the consistency with both measurements can be improved by increasing $|c_R|$ in Eq.~(\ref{ec:Zbb}). Since
$c_{L,R} = - X_{bb}^{L,R} + \frac{2}{3}s_W^2$ at the tree level, with $X = -2\,T_3$ for down-type quarks,
it can be easily seen that agreement with experimental data can be improved with a moderate mixing of $b_R$ with a heavy $B_R$ having weak isospin $T_3 > 0$, as it appears in the \by\ doublet only.\footnote{An interpretation of the observed top quark resonance as the lower member of a hypercharge $-5/6$ doublet~\cite{Chang:1999zc} is excluded by the several direct measurements of the top quark charge at the Tevatron~\cite{Abazov:2006vd,Aaltonen:2013sg} and the LHC~\cite{CMS:2012oua,Aad:2013uza}.}
Previous work~\cite{Choudhury:2001hs} has actually attempted to explain $A_\text{FB}^b$ via the simultaneous mixing with a \by\ doublet and a $B$ singlet. (See also~\cite{Batell:2012ca} for a fit in a custodial model with vector-like quarks, with implications for Higgs physics,~\cite{DaRold:2010as,Alvarez:2010js} for a composite Higgs model and~\cite{Bouchart:2008vp} for a model with a $B$ singlet and a new $Z'$ boson.) Mixing with the former increases $|c_R|$ in Eq.~(\ref{ec:Zbb}) and mixing with the latter decreases $|c_L|$, so that $R_b$ and $A_{FB}^b$ can be simultaneously fitted with two independent mixing parameters. Here we explore the simpler possibility of improving the agreement with the \by\ doublet only, fitting a single mixing parameter. Other possibilities involving large mixings and an overall change of sign in the couplings are not considered.

\begin{figure}[t]
\begin{center}
\begin{tabular}{cc}
\includegraphics[height=5.25cm]{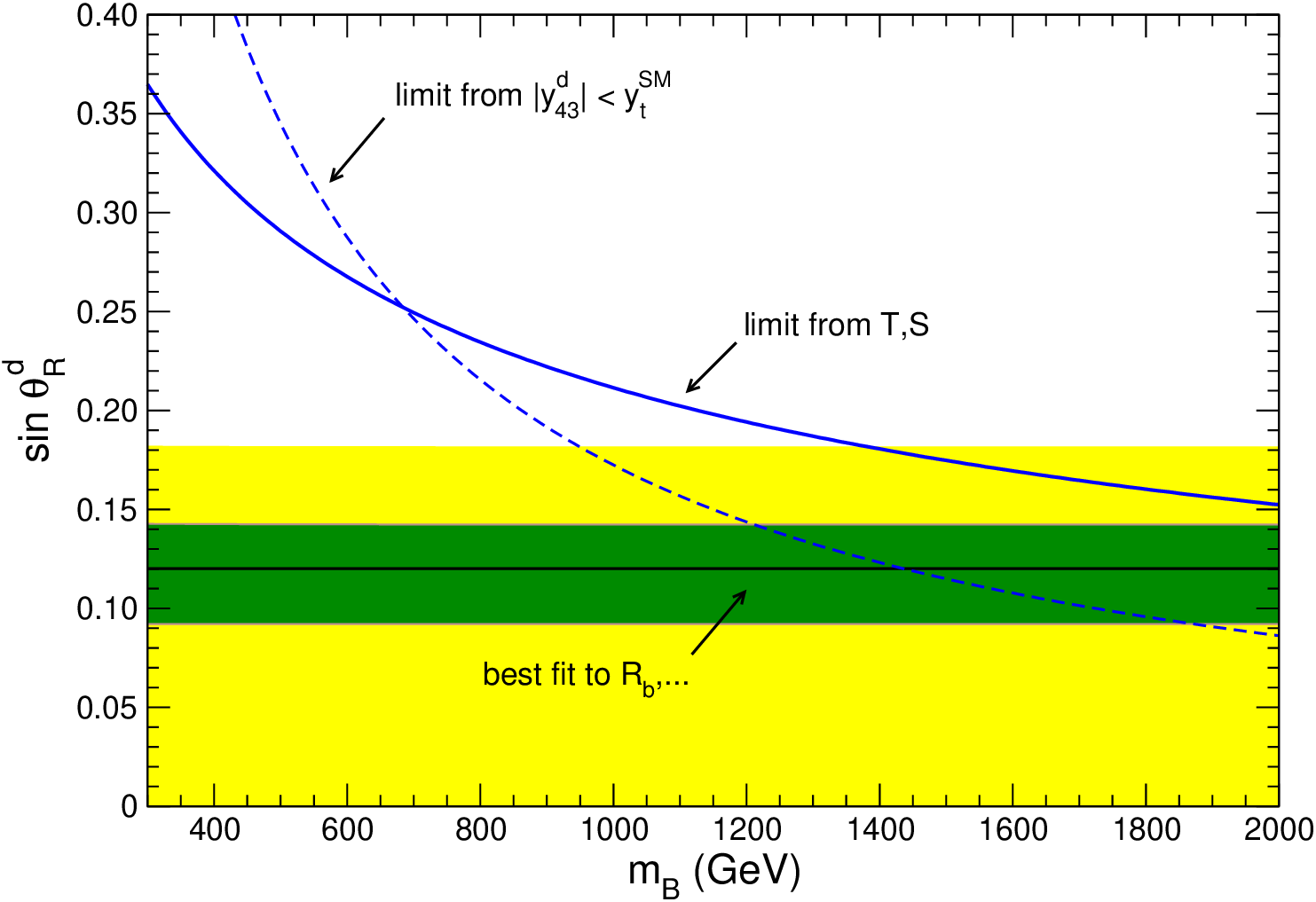} & \includegraphics[height=5.25cm]{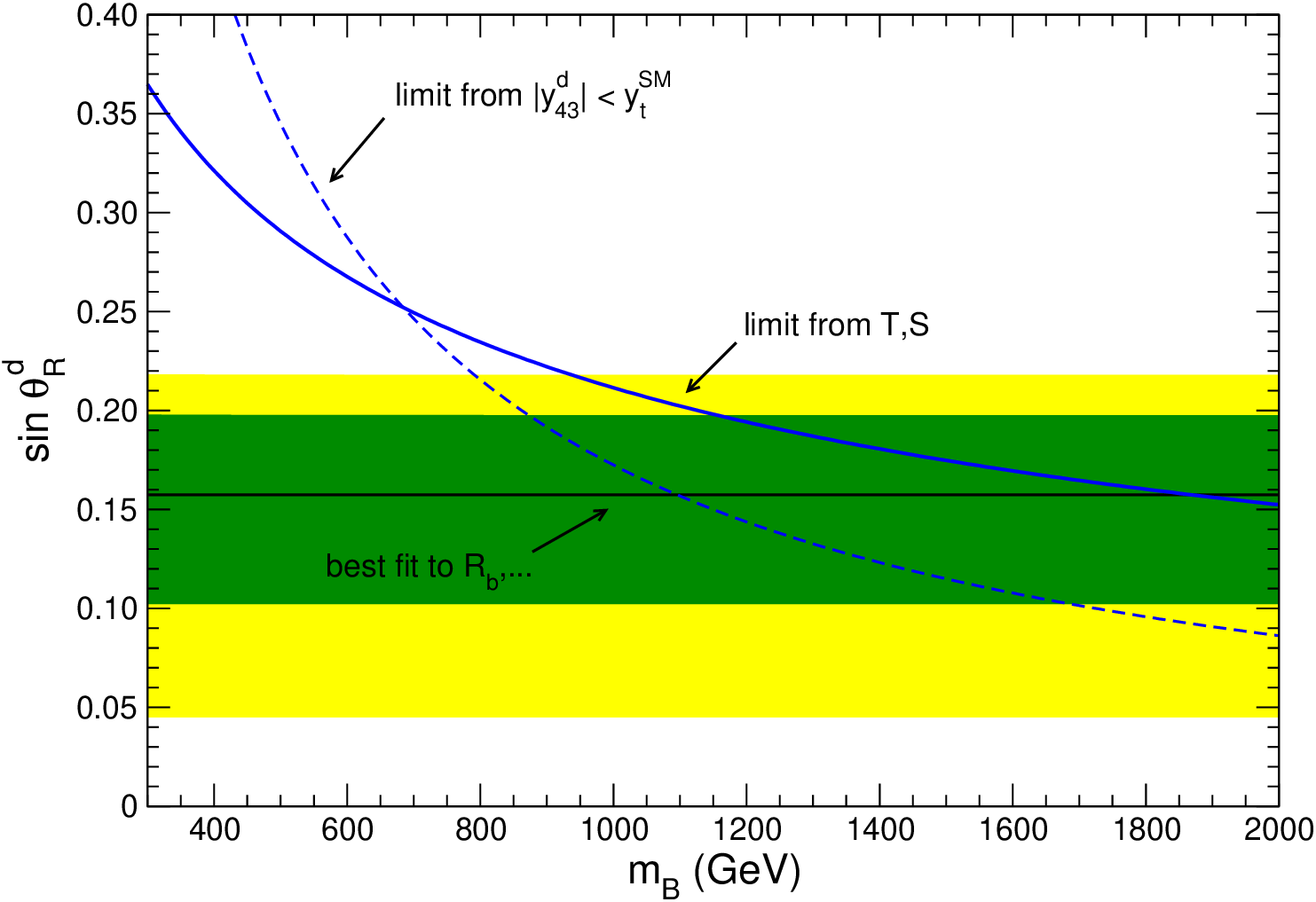}
\end{tabular}
\caption{Best-fit values of the mixing for the \by\ doublet for the two sets of predictions for $Z$ pole observables, from~\cite{Beringer:1900zz} (left) and~\cite{Baak:2012kk} (right). The upper constraints on mixing from oblique parameters are also included, as well as the line corresponding to an off-diagonal Yukawa equal to the SM top quark Yukawa.}
\label{fig:fit}
\end{center}
\end{figure}

We perform a fit to $Z$ pole observables in two different sets of predictions for the SM values. The first one~\cite{Beringer:1900zz} (hereafter called `fit 1') corresponds to the values of $R_b^\text{SM}$, $A_\text{FB}^{b,\text{SM}}$, $A_b^\text{SM}$ and $R_c^\text{SM}$ used in Section~\ref{sec:lim} to obtain upper limits on the mixing. The second scenario (`fit 2') corresponds to a new calculation of $R_b$ in the SM~\cite{Freitas:2012sy}, yielding the SM predictions from a fit~\cite{Baak:2012kk}
\begin{eqnarray}
R_b^\text{SM'} & = & 0.21474 \,, \notag \\
A_\text{FB}^{b,\text{SM'}} & = & 0.1032 \,, \notag \\
A_b^\text{SM'} & = & 0.93464 \,, \notag \\
R_c^\text{SM'} & = & 0.17223 \,.
\end{eqnarray}
Using the predictions in~\cite{Beringer:1900zz}, the best-fit value for the mixing is $\sin \theta_R = 0.12$, which reduces the $\chi^2$ from $\chi^2 = 7.37$ to $\chi^2 = 4.16$. Using the predictions in~\cite{Baak:2012kk} the best fit is obtained for $\sin \theta_R = 0.157$, which greatly improves the agreement with experimental data, from $\chi^2 = 10.97$ to $\chi^2 = 1.61$. The results are shown in Fig.~\ref{fig:fit}, together with the $1\sigma$ (green) and $2\sigma$ (yellow) bands.
Note that, for both fits, the preferred mixing with the doublet is smaller than the one obtained when a singlet $B$ is also included~\cite{Choudhury:2001hs,Batell:2012ca}.
We also point out that the results of the fit are independent of the heavy $B$ mass, since $m_B$ and $\theta_R^d$ are independent parameters and the corrections to $Zbb$ couplings only depend on $\theta_R^d$, see Eqs.~(\ref{ec:ll}) in Appendix~\ref{sec:a}. However, in order to get these mixings of order $0.1-0.2$, an off-diagonal Yukawa $y_{43}^d$ of order unity is needed in the mass matrix for the down sector, see Eq.~(\ref{ec:Lmass}). Imposing the loose requirement that this Yukawa is at most equal to the SM top quark Yukawa $y_t^\text{SM}$ (which is much larger than the bottom quark one), we obtain upper limits on the heavy mass, $m_B \lesssim 1.4$ TeV for fit 1, $m_B \lesssim 1.1$ TeV for fit 2. In addition, we have constraints from oblique corrections, which are more relevant for fit 2, $m_B \leq 1.9$ TeV.

\begin{figure}[t]
\begin{center}
\begin{tabular}{cc}
\includegraphics[height=5.4cm,clip=]{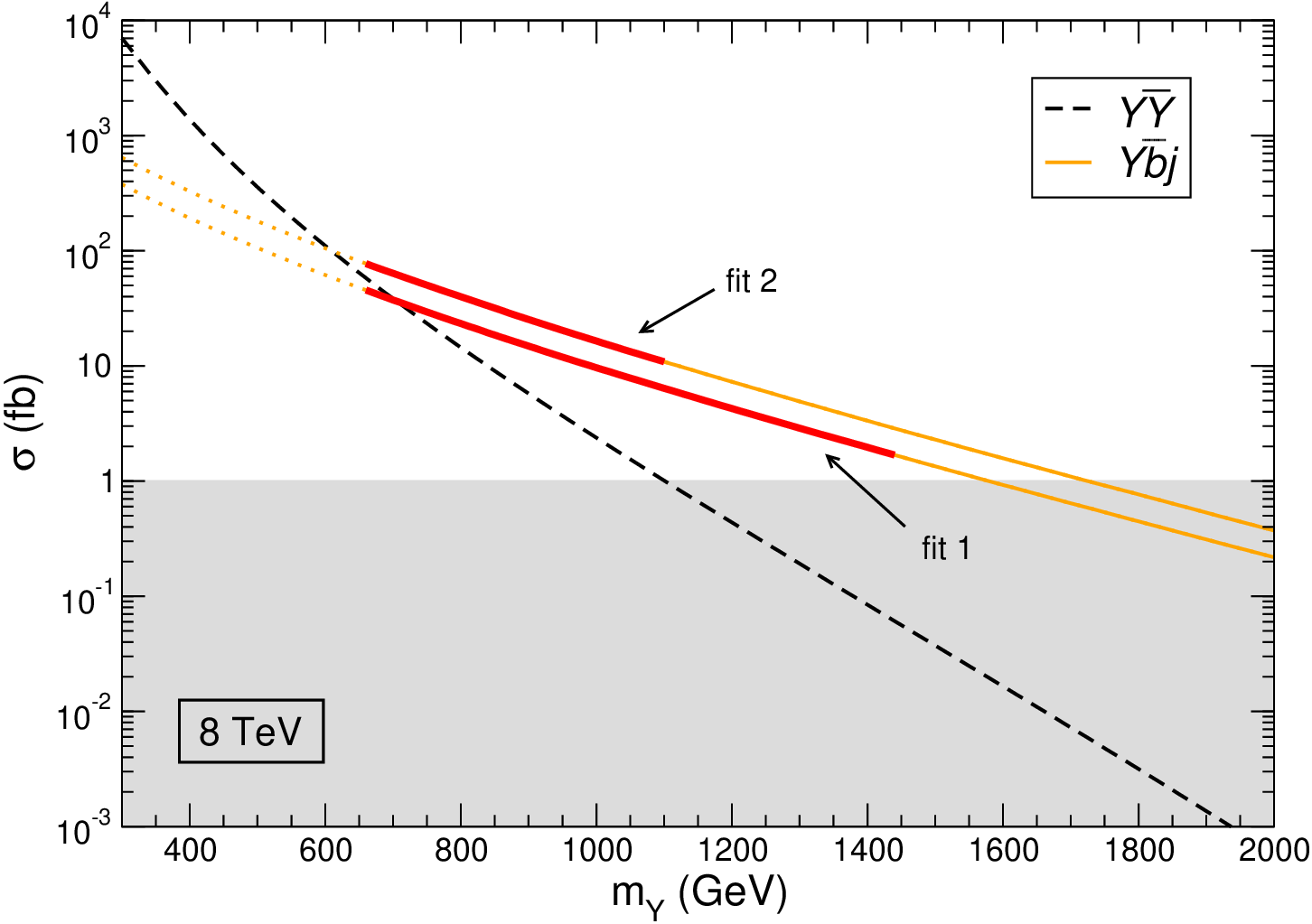} & \includegraphics[height=5.4cm,clip=]{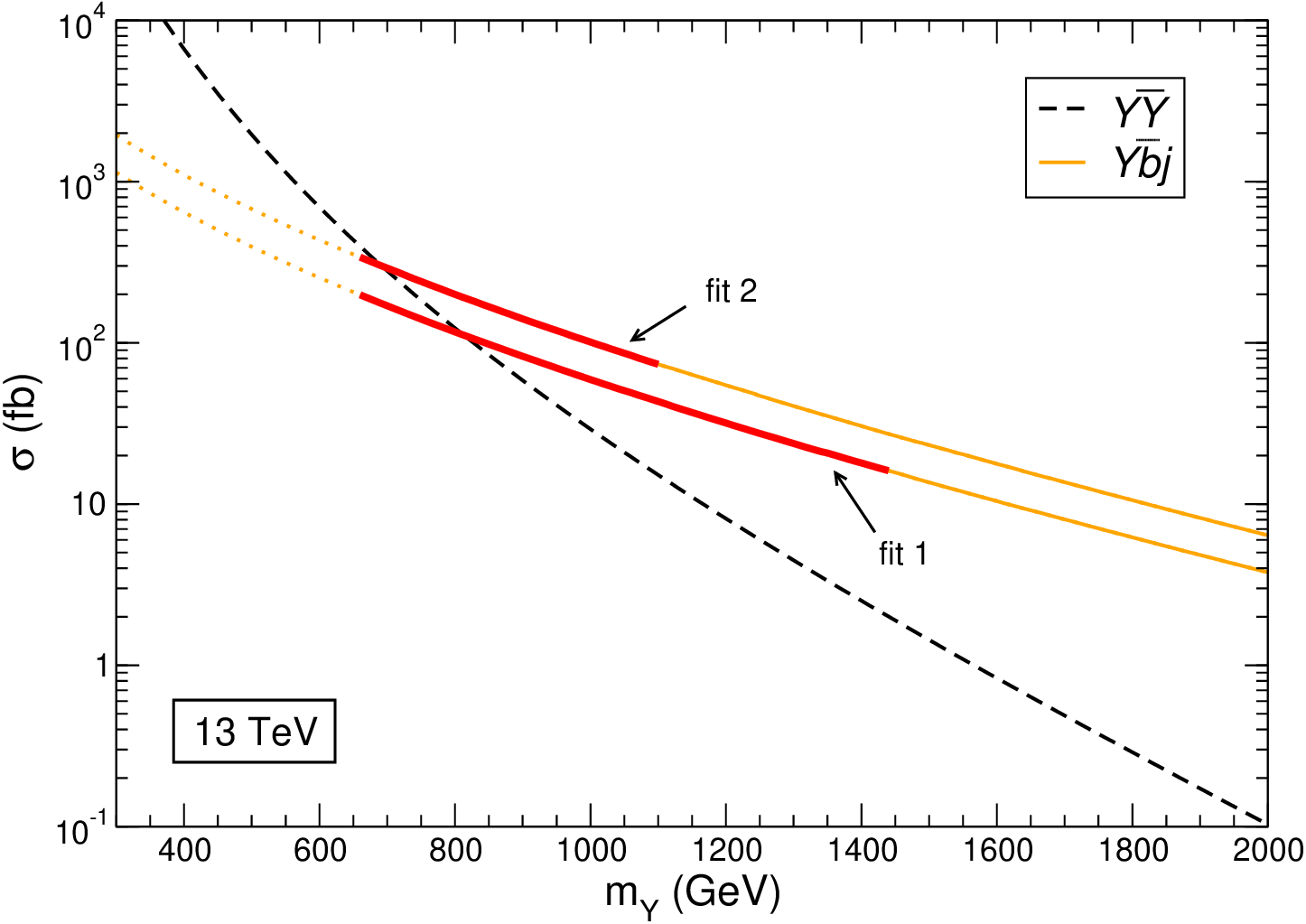}
\end{tabular}
\caption{Cross sections for $Y \bar b j$ production for the best-fit mixings at 8 TeV (left) and 13 TeV (right). The pair production cross section is shown for comparison.}
\label{fig:xsec-fit}
\end{center}
\end{figure}

New heavy quarks $Y$ with a mass of the order of the TeV and with a charged current coupling $V_{bY}^R \simeq 0.1$ are produced singly at large rates at the LHC, as seen in the previous section (see also~\cite{Kumar:2010vx}). We give in Fig.~\ref{fig:xsec-fit} the cross sections for the best-fit mixings corresponding to the two sets of $Z$ pole predictions.
The preferred mass range, between exclusion by direct searches and the upper limit from $|y_{43}^d|< y_t^\text{SM}$, is displayed by a thicker red line. It is therefore apparent that, if a \by\ doublet is the responsible for the deviation in the $Z \to b \bar b$ measurements, the new quark $Y$ should be seen at the LHC, perhaps already at the 8 TeV run.


\section{Summary}
\label{sec:sum}

New vector-like quarks can naturally have masses above the electroweak symmetry breaking scale. They are being searched for at the LHC, with lower limits on their masses in the range $600-800$ GeV, at present. These limits typically imply a small mixing with the SM quarks, in order to fulfil indirect constraints from oblique corrections and $Z \to b \bar b$ data. We have explicitly obtained these constraints for minimal SM extensions with one vector-like quark multiplet, with dominant coupling to the third generation. In the up quark sector this assumption is based on theoretical arguments and also on experimental data, which sets much stronger constraints on mixing with the first two generations than with the top quark. In the down sector the assumption of dominant mixing with the $b$ quark is motivated by the usual hierarchy of Yukawa couplings, but is not an experimental requisite. On the other hand, the simplification of studying one vector-like multiplet at a time is expected to lead to conservative bounds for generic theories with several quark multiplets, but it can break down in the case of specific models that make use of cancellations between the contributions of different multiplets to electroweak observables. We have also checked that Higgs data do not impose further constraints on these minimal scenarios with vector-like quarks.

A first outcome of our analysis, very relevant for heavy quark searches at the LHC, concerns the decay of the heavy quarks.
It is often mentioned in the literature that the mass splitting for vector-like multiplets should be small, since it is induced by electroweak breaking. We have given explicit upper bounds on the size of this effect. The mass splittings are at most of the order of several GeV, so the decays from one heavy quark to another one are suppressed. The only allowed decays are then into a SM quark $t,b$ plus a $W$, $Z$ or Higgs boson. We give predictions for the branching ratios of $T$, $B$ quarks in all multiplet representations, taking into account the suppressed mixings and mass effects.

Single production of heavy quarks at the LHC, which is less suppressed by parton distribution functions but is proportional to the mixing squared, becomes more relevant as lower limits on their mass grow. In this respect, we have obtained the maximum single production cross sections for the multiplets $T$, \xt, \tb, \by, where the mixing with the SM quarks can be largest, identifying the most promising processes: $Y \bar b j$ and $B \bar b j$ for the \by\ doublet and $T \bar b j$ for the $T$ singlet and the \tb\ doublet. Our calculations of the maximal cross sections allowed by indirect constraints can be used as a guide for LHC searches in the standard single production channels. We have not considered here other model-dependent production mechanisms, which may be available in the presence of additional new particles, such as heavy gluons~\cite{Dobrescu:2009vz,Vignaroli:2011ik,Barcelo:2011wu,Chala:2013ega,Redi:2013eaa}.

With the constraints we have obtained, we find that the deviations in top couplings induced by mixing with vector-like quarks are small in general. Only for the $Ztt$ vertex these deviations might be visible in $e^+ e^- \to t \bar t$ at a future ILC, provided the systematic uncertainties on asymmetry measurements are kept very small, which would a challenge for the data analysis. However, on the bottom side, we have addressed the possibility of explaining the anomalous asymmetry $A_\text{FB}^b$ in $Z \to b \bar b$ at LEP by the mixing of the $b$ quark with a \by\ doublet (one of the ``beautiful mirrors'' in~\cite{Choudhury:2001hs}). We have found that such mixing improves significantly the $\chi^2$ of the fit of the relevant observables, especially when one takes into account the recent calculation of two-loop electroweak corrections to $R_b$~\cite{Freitas:2012sy}: the improvement is then from $\chi^2 = 10.97$ in the SM (with 4 degrees of freedom) to a best-fit value $\chi^2=1.61$ (with 3 degrees of freedom). This best fit is independent of the heavy quark masses, but we have given reasons to expect them in the range around 1 TeV. In that case, the new quark $Y$ of charge $-4/3$ might be seen in current 8 TeV data, and would definitely be observed in the second LHC run at 13 TeV. In addition, this explanation of the $A_\text{FB}^b$ anomaly predicts deviations in the $H \to b \bar b$, $H \to gg$ branching ratios that would be visible at the ILC at the $2\sigma$ level.
The \by\ doublet constitutes a unique example of a model that can improve the agreement of the predictions for $Z \to b \bar b$ with experimental data by fitting a single mixing parameter,\footnote{An explanation of this anomaly with a $Z^\prime$ in terms of a single parameter is also possible, but it involves large, possibly non-perturbative, couplings. The couplings can be made smaller including additional new particles~\cite{delAguila:2010mx}, but this complicates the model.} and predicting at the same time a likely range for the mass of the a new particle, which can be probed at the LHC.

Before finishing, we would like to point out to the reader, once more, that all the quantitative statements made in this paper regarding the possible effects of mixing in top couplings, single production cross sections, etc. may be relaxed if more than one vector-like multiplet (or other kind of new physics) is introduced, such that the cancellations between different contributions hide the indirect effects of the new particles. For instance, models can be constructed along the lines of~\cite{Atre:2011ae} to allow for large $\mathcal{O}(0.5)$ mixings, leading to large deviations in top couplings and large cross sections for the suppressed processes $X \bar t j$, $T \bar t j$.  Because substantial cancellations are non-trivial, the compatibility with LEP data of large mixings must be tested on a model by model basis. For other kinds of SM extensions with generic parameters, on the other hand, we expect that all the results we have derived for the minimal scenarios hold, at least qualitatively. 

To conclude, in this paper we have addressed a variety of direct and indirect effects resulting from the mixing of vector-like quarks with the third generation. This work complements the study of heavy quark pair production in~\cite{AguilarSaavedra:2009es} by addressing single production, by refining predictions for the heavy quark decays, and by the inclusion of vector-like triplets into the game. Together, these two works provide a comprehensive guide for the LHC phenomenology of minimal extensions of the SM with vector-like quarks.

\section*{Acknowledgements}
We thank J. Santiago for useful discussions.
This work has been supported by MICINN by projects FPA2006-05294, FPA2010-17915, FPA2010-22163-C02-01 and by the Consolider-Ingenio 2010 Program under grant MultiDark CSD2009-00064; by
Junta de Andaluc\'{\i}a (FQM 101, FQM 03048 and FQM 6552) and by Funda\c c\~ao
para a Ci\^encia e Tecnologia~(FCT) project CERN/ FP/123619/2011. The work of R.B. was supported by the Spanish Consejo Superior de Investigaciones Cientificas (CSIC).

\appendix

\section{Lagrangian}
\label{sec:a}

The Lagrangian of the SM quarks $t,b$ modified by the mixing with vector-like quarks reads
\begin{eqnarray}
\mathcal{L}_W & = & -\frac{g}{\sqrt 2} \bar t \gm \left( V_{tb}^L P_L + V_{tb}^R P_R \right) b \Wm^+ +\text{H.c.} \,, \notag \\ 
\mathcal{L}_Z & = & -\frac{g}{2 c_W} \bar t \gm \left( X_{tt}^L P_L + X_{tt}^R P_R - 2 Q_t s_W^2 \right) t \Zm \notag \\
& & -\frac{g}{2 c_W} \bar b \gm \left( -X_{bb}^L P_L - X_{bb}^R P_R - 2 Q_b s_W^2 \right) b \Zm \,, \notag \\
\mathcal{L}_H & = & -\frac{g m_t}{2M_W} Y_{tt} \bar t t H -\frac{g m_b}{2M_W} Y_{bb} \bar b b H \,,
\label{ec:ll}
\end{eqnarray}
plus the interactions with the gluon and photon that remain the same as in the SM. The charged current mixings $V_{tb}^{L,R}$ for all multiplets are given in Table~\ref{tab:llW}, the neutral ones $X_{tt}^{L,R}$, $X_{bb}^{L,R}$ in Table~\ref{tab:llZ} and the Higgs couplings $Y_{tt}$, $Y_{bb}$ in Table~\ref{tab:llH}.

\begin{table}[htb]
\begin{center}
\begin{tabular}{c|cc}
& $V_{tb}^L$ & $V_{tb}^R$ 
\\ \hline
\ts & $\clx$ & 0
\\
\bs & $\clx$ & 0
\\
\xt & $\clx$ & 0
\\
\tb & $\clu \cld + \slu \sld e^{i(\phi_u-\phi_d)}$ & $\sru \srd e^{i(\phi_u-\phi_d)}$
\\
\by & $\clx$ & 0
\\
\xtb & $\clu \cld + \sqt \slu \sld$ & $\sqt \sru \srd$
\\
\tby & $\clu \cld + \sqt \slu \sld$ & $\sqt \sru \srd$
\end{tabular}
\caption{Light-light couplings to the $W$ boson.}
\label{tab:llW}
\end{center}
\end{table}

\begin{table}[htb]
\begin{center}
\begin{tabular}{c|cccc}
& $X_{tt}^L$ & $X_{tt}^R$ & $X_{bb}^L$ & $X_{bb}^R$ 
\\ \hline
\ts & $\clx^2$ & 0 & 1 & 0
\\
\bs & 1 & 0 & $\clx^2$ & 0
\\
\xt & $\clx^2-\slx^2$ & $-\srx^2$ & 1 & 0
\\
\tb & 1 & $(\sru)^2$ & 1 & $(\srd)^2$
\\
\by & 1 & 0 & $\clx^2-\slx^2$ & $-\slx^2$
\\
\xtb & $(\clu)^2$ & 0 & $1+(\sld)^2$ & $2(\srd)^2$
\\
\tby & $1+(\slu)^2$ & $2(\sru)^2$ & $(\cld)^2$ & 0
\end{tabular}
\caption{Light-light couplings to the $Z$ boson.}
\label{tab:llZ}
\end{center}
\end{table}

\begin{table}[htb]
\begin{center}
\begin{tabular}{c|cc}
& $Y_{tt}$ & $Y_{bb}$  
\\ \hline
\ts & $\clx^2$ & $1$
\\
\bs & $1$ & $\clx^2$
\\
\xt & $\crx^2$ & $1$
\\
\tb & $(\cru)^2$ & $(\crd)^2$
\\
\by & $1$ & $\crx^2$ 
\\
\xtb & $(\clu)^2$ & $(\cld)^2$
\\
\tby & $(\clu)^2$ & $(\cld)^2$
\end{tabular}
\caption{Light-light couplings to the Higgs boson.}
\label{tab:llH}
\end{center}
\end{table}


The Lagrangian for the heavy quarks $Q,Q'=X,T,B,Y$ follow a similar notation,
\begin{eqnarray}
\mathcal{L}_W & = & -\frac{g}{\sqrt 2} \bar Q \gm \left( V_{QQ'}^L P_L + V_{QQ'}^R P_R \right) Q' \Wm^+ +\text{H.c.} \,, \notag \\ 
\mathcal{L}_Z & = & -\frac{g}{2 c_W} \bar Q \gm \left( \pm X_{QQ}^L P_L \pm X_{QQ}^R P_R - 2 Q_Q s_W^2 \right) Q \Zm \,, \notag \\
\mathcal{L}_H & = & -\frac{g m_Q}{2M_W} Y_{QQ} \bar Q Q H \,,
\label{ec:HH}
\end{eqnarray}
with the plus (minus) sign in the $Z$ term for $X,T$ ($B,Y$). Charged current mixings $V_{QQ'}^{L,R}$ are given in Table~\ref{tab:hhW}, neutral ones $X_{QQ}^{L,R}$ in Table~\ref{tab:hhZ} and the Higgs couplings $Y_{QQ}$ in Table~\ref{tab:hhH}. The electromagnetic interactions are determined by the quark charge, and the strong interactions are the same as for any other quark.

\begin{table}[htb]
\begin{center}
\begin{tabular}{c|cccccc}
& $V_{XT}^L$ & $V_{XT}^R$ & $V_{TB}^L$ & $V_{TB}^R$ & $V_{BY}^L$ & $V_{BY}^R$
\\ \hline
\xt & $\clx$ & $\crx$ & -- & -- & -- & --
\\
\tb & -- & -- & $\clu \cld + \slu \sld e^{-i (\phi_u-\phi_d)}$ & $\cru \crd$ & -- & --
\\
\by & -- & -- & -- & -- & $\clx$ & $\crx$
\\
\xtb & $\sqt \clu$ & $\sqt \cru$ & $\slu \sld + \sqt \clu \cld$ & $\sqt \cru \crd$ & -- & --
\\
\tby & -- & -- & $\slu \sld + \sqt \clu \cld$ & $\sqt \cru \crd$ & $\sqt \cld$ & $\sqt \crd$
\end{tabular}
\caption{Heavy-heavy couplings to the $W$ boson.}
\label{tab:hhW}
\end{center}
\end{table}

\begin{table}[htb]
\begin{center}
\begin{tabular}{c|cccccccc}
& $X_{XX}^L$ & $X_{XX}^R$ & $X_{TT}^L$ & $X_{TT}^R$ & $X_{BB}^L$ & $X_{BB}^R$ & $X_{YY}^L$ & $X_{YY}^R$
\\ \hline
\ts & -- & -- & $\slx^2$ & 0 & -- & -- & -- & --
\\
\bs & -- & -- & -- & -- & $\slx^2$ & 0 & -- & --
\\
\xt & 1 & 1 & $\slx^2 - \clx^2$ & $-\crx^2$ & -- & -- & -- & --
\\
\tb & -- & -- & 1 & $(\cru)^2$ & 1 & $(\crd)^2$ & -- & --
\\
\by & -- & -- & -- & -- & $\slx^2 - \clx^2$ & $-\crx^2$ & 1 & 1
\\
\xtb & 2 & 2 & $(\slu)^2$ & 0 & $1+(\cld)^2$ & $2 (\crd)^2$ & -- & --
\\
\tby & -- & -- & $1+(\clu)^2$ & $2 (\cru)^2$ & $(\sld)^2$ & 0 & 2 & 2
\end{tabular}
\caption{Heavy-heavy couplings to the $Z$ boson.}
\label{tab:hhZ}
\end{center}
\end{table}

\begin{table}[htb]
\begin{center}
\begin{tabular}{c|cccc}
& $Y_{XX}$ & $Y_{TT}$ & $Y_{BB}$ & $Y_{YY}$ 
\\ \hline
\ts & -- & $\slx^2$ & -- & --
\\
\bs & -- & -- & $\slx^2$ & --
\\
\xt & 0 & $\srx^2$ & -- & --
\\
\tb & -- & $(\sru)^2$ & $(\srd)^2$ & --
\\
\by & -- & -- & $\srx^2$ & 0
\\
\xtb & 0 & $(\slu)^2$ & $(\sld)^2$ & --
\\
\tby & -- & $(\slu)^2$ & $(\sld)^2$ & 0
\end{tabular}
\caption{Heavy-heavy couplings to the Higgs boson.}
\label{tab:hhH}
\end{center}
\end{table}


Finally, the terms involving a heavy ($Q$) and a light ($q$) quark are
\begin{eqnarray}
\mathcal{L}_W & = & -\frac{g}{\sqrt 2} \bar Q \gm \left( V_{Qq}^L P_L + V_{Qq}^R P_R \right) q \Wm^+ +\text{H.c.} \notag \\
& & -\frac{g}{\sqrt 2} \bar q \gm \left( V_{qQ}^L P_L + V_{qQq}^R P_R \right) Q \Wm^+ +\text{H.c.} \,, \notag \\ 
\mathcal{L}_Z & = & -\frac{g}{2 c_W} \bar q \gm \left( \pm X_{qQ}^L P_L \pm X_{qQ}^R P_R \right) Q \Zm +\text{H.c.} \,, \notag \\
\mathcal{L}_H & = & -\frac{g m_Q}{2M_W} \bar q \left( Y_{qQ}^L P_L +  Y_{qQ}^R P_R \right) Q H +\text{H.c.} \,,
\end{eqnarray}

\begin{table}[htb]
\begin{center}
\begin{tabular}{c|cccc}
& $V_{Xt}^L$ & $V_{Xt}^R$ & $V_{Tb}^L$ & $V_{Tb}^R$ 
\\ \hline
\ts & -- & -- & $\slx e^{-i \phi}$ & 0 
\\
\xt & $-\slx e^{-i\phi}$ & $-\srx e^{-i\phi}$ & $\slx e^{-i\phi}$ & 0 
\\
\tb & -- & -- & $\slu \cld e^{-i \phi_u} - \clu \sld e^{-i \phi_d}$ & $-\cru \srd e^{-i \phi_d}$
\\
\xtb & $-\sqt \slu e^{-i\phi}$ & $-\sqt \sru e^{-i\phi}$ & $(\slu \cld - \sqt \clu \sld) e^{-i\phi}$ & $-\sqt \cru \srd e^{-i\phi}$
\\
\tby & -- & -- & $(\slu \cld - \sqt \clu \sld) e^{-i\phi}$ & $-\sqt \cru \srd e^{-i\phi}$
\end{tabular}
\caption{Heavy-light couplings to the $W$ boson.}
\label{tab:WHl}
\end{center}
\end{table}

\begin{table}[htb]
\begin{center}
\begin{tabular}{c|cccc}
& $V_{tB}^L$ & $V_{tB}^R$ & $V_{bY}^L$ & $V_{bY}^R$
\\ \hline
\bs & $\slx e^{i\phi}$ & 0 & -- & --
\\
\tb & $\clu \sld e^{i \phi_d}-\slu \cld e^{i \phi_u}$ & $-\sru \crd e^{i \phi_u}$ & -- & --
\\
\by & $\slx e^{i\phi}$ & 0 & $-\slx e^{i\phi}$ & $-\srx e^{i\phi}$
\\
\xtb & $(\clu \sld - \sqt \slu \cld) e^{i\phi}$ & $-\sqt \sru \crd e^{i\phi}$ & -- & --
\\
\tby & $(\clu \sld - \sqt \slu \cld)e^{i\phi}$ & $-\sqt \sru \crd e^{i\phi}$ & $-\sqt \sld e^{i\phi}$ & $-\sqt \srd e^{i\phi}$
\end{tabular}
\caption{Light-heavy couplings to the $W$ boson.}
\label{tab:WlH}
\end{center}
\end{table}

\begin{table}[htb]
\begin{center}
\begin{tabular}{c|cccc}
& $X_{tT}^L$ & $X_{tT}^R$ & $X_{bB}^L$ & $X_{bB}^R$ 
\\ \hline
\ts & $\slx \clx e^{i\phi}$ & 0 & -- & --
\\
\bs & -- & -- & $\slx \clx e^{i\phi}$ & 0
\\
\xt & $2 \slx \clx e^{i\phi}$ & $\srx \crx e^{i\phi}$ & -- & --
\\
\tb & 0 & $-\sru \cru e^{i\phi_u}$ & 0 & $-\srd \crd e^{i\phi_d}$
\\
\by & -- & -- & $2\slx \clx e^{i\phi}$ & $\srx \crx e^{i\phi}$ 
\\
\xtb & $\slu \clu e^{i\phi}$ & 0 & $-\sld \cld e^{i\phi}$ & $-2\srd \crd e^{i\phi}$
\\
\tby & $-\slu \clu e^{i\phi}$ & $-2\sru \cru e^{i\phi}$ & $\sld \cld e^{i\phi}$ & 0 
\end{tabular}
\caption{Light-heavy couplings to the $Z$ boson.}
\label{tab:ZlH}
\end{center}
\end{table}

\clearpage

\begin{table}[htb]
\begin{center}
\begin{tabular}{c|cccc}
& $Y_{tT}^L$ & $Y_{tT}^R$ & $Y_{bB}^L$ & $Y_{bB}^R$  
\\ \hline
\ts & $\frac{m_t}{m_T} \slx \clx e^{i\phi}$ & $\slx \clx e^{i\phi}$ & -- & --
\\
\bs & -- & -- & $\frac{m_b}{m_B} \slx \clx e^{i\phi}$ & $\slx \clx e^{i\phi}$
\\
\xt & $\srx \crx e^{i\phi}$ & $\frac{m_t}{m_T} \srx \crx e^{i\phi}$ & -- & --
\\
\tb & $\sru \cru e^{i\phi_u}$ & $\frac{m_t}{m_T} \sru \cru e^{i\phi_u}$ & $\srd \crd e^{i\phi_d}$ & $\frac{m_b}{m_B} \srd \crd e^{i\phi_d}$
\\
\by & -- & -- & $\srx \crx e^{i\phi}$ & $\frac{m_b}{m_B} \srx \crx e^{i\phi}$ 
\\
\xtb & $\frac{m_t}{m_T} \slu \clu e^{i\phi}$  & $\slu \clu e^{i\phi}$ & $\frac{m_b}{m_B} \sld \cld e^{i\phi}$ & $\sld \cld e^{i\phi}$
\\
\tby & $\frac{m_t}{m_T} \slu \clu e^{i\phi}$ & $\slu \clu e^{i\phi}$ & $\frac{m_b}{m_B} \sld \cld e^{i\phi}$ & $\sld \cld e^{i\phi}$
\end{tabular}
\caption{Light-heavy couplings to the Higgs boson.}
\label{tab:HlH}
\end{center}
\end{table}

\section{Heavy quark decay widths}
\label{sec:b}

The total widths for the decay of $X$, $Y$ quarks are (see also~\cite{Atre:2011ae})
\begin{align}
\Gamma(X \to W^+ t) & = \frac{g^2}{64 \pi}  \frac{m_X}{M_W^2} \lambda(m_X,m_t,M_W)^{1/2} \left\{
(|V_{Xt}^L|^2+|V_{Xt}^R|^2)  \right. \nonumber \\
  & \left. \times \left[ 1+r_W^2-2 r_t^2
  -2 r_W^4  + r_t^4 +r_W^2 r_t^2
  \right]  -12 r_W^2 r_t \RE V_{Xt}^L V_{Xt}^{R*} \right\} \,, \notag \\
\Gamma(Y \to W^- b) & = \frac{g^2}{64 \pi}  \frac{m_T}{M_W^2} \lambda(m_Y,m_b,M_W)^{1/2} \left\{
(|V_{bY}^L|^2+|V_{bY}^R|^2)  \right. \nonumber \\
  & \left. \times \left[ 1+r_W^2-2 r_b^2
  -2 r_W^4  + r_b^4 +r_W^2 r_b^2
  \right]  -12 r_W^2 r_b \RE V_{bY}^L V_{bY}^{R*} \right\}
\,.
\label{ec:GammaXY}
\end{align}
with $r_x \equiv m_x / m_Q$, where $x=t,b,W,Z,H$ and $Q$ is the heavy quark, and
\begin{equation}
\lambda(x,y,z) \equiv (x^4 + y^4 + z^4 - 2 x^2 y^2 
- 2 x^2 z^2 - 2 y^2 z^2) \,.
\end{equation}%
The charged current mixings $V$ are given in Tables~\ref{tab:WlH} and~\ref{tab:WHl} of Appendix~\ref{sec:a}.
The partial widths for $T$ decays, including all possible mixing terms, are
\begin{align}
\Gamma(T \to W^+ b) & = \frac{g^2}{64 \pi}  \frac{m_T}{M_W^2} \lambda(m_T,m_b,M_W)^{1/2} \left\{
(|V_{Tb}^L|^2+|V_{Tb}^R|^2)  \right. \nonumber \\
  & \left. \times \left[ 1+r_W^2-2 r_b^2 -2 r_W^4  + r_b^4 +r_W^2 r_b^2
  \right]  -12 r_W^2 r_b \RE V_{Tb}^L V_{Tb}^{R*} \right\}
\,, \nonumber \\
\Gamma(T \to Z t) & = \frac{g}{128 \pi c_W^2}  \frac{m_T}{M_Z^2} \lambda(m_T,m_t,M_Z)^{1/2}
\left\{ (|X_{tT}^L|^2 + |X_{tT}^R|^2) \right. \nonumber \\
  & \left. \times  \left[ 1 + r_Z^2 - 2  r_t^2 - 2  r_Z^4  + r_t^4
  + r_Z^2 r_t^2 \right]  -12 r_Z^2 r_t \RE X_{tT}^L X_{tT}^{R*}  \right\} \,, \nonumber \\
\Gamma(T \to H t) & = \frac{g^2}{128 \pi}
 \frac{m_T}{M_W^2} \lambda(m_T,m_t,M_H)^{1/2} |Y_{tT}|^2 \left[ 1 + 6 r_t^2 - r_H^2 
  + r_t^4 - r_t^2 r_H^2 \right] \,,
\label{ec:GammaT}
\end{align}
and for the $B$ quark they are completely analogous,
\begin{align}
\Gamma(B \to W^- t) & = \frac{g^2}{64 \pi}  \frac{m_B}{M_W^2} \lambda(m_B,m_t,M_W)^{1/2} \left\{
(|V_{tB}^L|^2+|V_{tB}^R|^2)  \right. \nonumber \\
  & \left. \times \left[ 1+r_W^2-2 r_t^2  -2 r_W^4  + r_t^4 +r_W^2 r_t^2
  \right]  -12 r_W^2 r_t \RE V_{tB}^L V_{tB}^{R*} \right\}
\,, \nonumber \\
\Gamma(B \to Z b) & = \frac{g}{128 \pi c_W^2}  \frac{m_B}{M_Z^2} \lambda(m_B,m_b,M_Z)^{1/2}
\left\{ (|X_{bB}^L|^2 + |X_{bB}^R|^2) \right. \nonumber \\
  & \left. \times  \left[ 1 + r_Z^2 - 2  r_b^2 - 2  r_Z^4  + r_b^4
  + r_Z^2 r_b^2 \right]  -12 r_Z^2 r_b \RE X_{bB}^L X_{bB}^{R*}  \right\} \,, \nonumber \\
\Gamma(B \to H b) & = \frac{g^2}{128 \pi}
 \frac{m_B}{M_W^2} \lambda(m_B,m_b,M_H)^{1/2} |Y_{bB}|^2 \left[ 1 + 6 r_b^2 - r_H^2 
 + r_b^4 - r_b^2 r_H^2 \right] \,.
\label{ec:GammaB}
\end{align}
The neutral current mixings $X$ are given in Table~\ref{tab:ZlH} of Appendix~\ref{sec:a}. In the partial widths to Higgs final states, $Y_{qQ}$ refers to the dominant light-heavy Yukawa coupling in Table~\ref{tab:HlH}, that is,
$Y_{qQ}^R$ for the singlets and triplets, $Y_{qQ}^L$ for the doublets.

\end{document}